\documentclass[sigconf, nonacm]{acmart} 

\usepackage{subcaption}

\usepackage{array}
\newcolumntype{L}[1]{>{\raggedright\let\newline\\\arraybackslash\hspace{0pt}}m{#1}}
\newcolumntype{C}[1]{>{\centering\let\newline\\\arraybackslash\hspace{0pt}}m{#1}}
\newcolumntype{R}[1]{>{\raggedleft\let\newline\\\arraybackslash\hspace{0pt}}m{#1}}

\usepackage{enumitem}

\begin{document}
\title{Water Simulation and Rendering from a Still Photograph}

\author{Ryusuke Sugimoto}
\orcid{0000-0001-5894-0423}
\affiliation{
 \institution{University of Waterloo}
 \city{Waterloo}
 \state{ON}
 \country{Canada}}
\email{rsugimot@uwaterloo.ca}

\author{Mingming He}
\orcid{0000-0002-9982-7934}
\affiliation{%
 \institution{Netflix}
 \city{Los Angeles}
 \state{CA}
 \country{USA}}
\email{hmm.lillian@gmail.com}

\author{Jing Liao}
\orcid{0000-0001-7014-5377}
\affiliation{
 \institution{City University\\of Hong Kong}
 \city{Kowloon}
 \country{Hong Kong}
 }
\email{jingliao@cityu.edu.hk}

\author{Pedro V. Sander}
\orcid{0000-0002-0435-9833}
\affiliation{
 \institution{The Hong Kong University\\of Science and Technology}
 \city{Kowloon}
 \country{Hong Kong}
 }
\email{psander@cse.ust.hk}

\begin{teaserfigure}
\centering
\includegraphics[width=1.\linewidth]{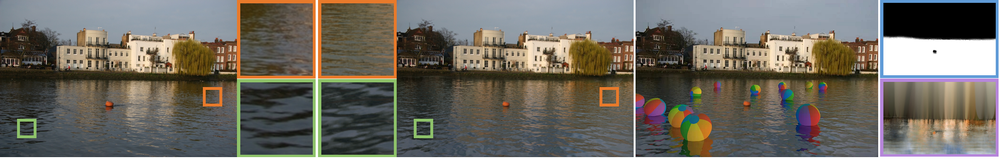}
\begin{tabular}{C{0.23\textwidth}C{0.1\textwidth}C{0.23\textwidth}C{0.23\textwidth}C{0.1\textwidth}}
\small{\textsf{(a) Input}} & & \small{\textsf{(b) Rendered}} & \small{\textsf{(c) Edited}} & \small{\textsf{(d) Msk.\&Ref.}}\\
\end{tabular}
\caption{Given a single input image (a), our approach estimates the parameters, segmentation mask and reflection texture (d) needed to predict and render a realistic animated water surface (b), further enabling interactive editing with the water by placing synthetic objects (\textit{e.g.} beach balls) on the surface (c). Input image: - Paul -/Flickr.}
\label{fig:teaser}
\end{teaserfigure}

\begin{abstract}
We propose an approach to simulate and render realistic water animation from a single still input photograph. We first segment the water surface, estimate rendering parameters, and compute water reflection textures with a combination of neural networks and traditional optimization techniques. Then we propose an image-based screen space local reflection model to render the water surface overlaid on the input image and generate real-time water animation. Our approach creates realistic results with no user intervention for a wide variety of natural scenes containing large bodies of water with different lighting and water surface conditions. Since our method provides a 3D representation of the water surface, it naturally enables direct editing of water parameters and also supports interactive applications like adding synthetic objects to the scene.
\end{abstract}

\ccsdesc[500]{Computing methodologies~Computer graphics}
\ccsdesc[300]{Computing methodologies~Image-based rendering}
\ccsdesc[300]{Computing methodologies~Computational photography}

\keywords{single-image animation generation, texture prediction, neural networks, optimization, screen-space reflection}

\maketitle

\section{Introduction and Related Work}

Simulating and rendering water is an extensively studied problem in traditional computer graphics~\cite{Darles2011,Bridson2007}. It involves simulating the water dynamics using parametric models~\cite{Tessendorf2001,DBLP:conf/siggraph/Max81} and generating realistic shading that takes into account various optical properties of the water surface~\cite{Tessendorf2001}. Empirical parametric models can generate visually appealing water geometry and appearance and allow the result to extend arbitrarily in space and time. However, they usually rely on a large number of parameters to model the water surface dynamics (\textit{e.g.} wind speed, wind direction, \textit{etc.}) and lighting conditions. This often requires much time and manual effort to tune the parameters to achieve the desirable effect.

\begin{figure*}[ht]
    \includegraphics[trim=220 80 140 40, clip,width=0.98\hsize]{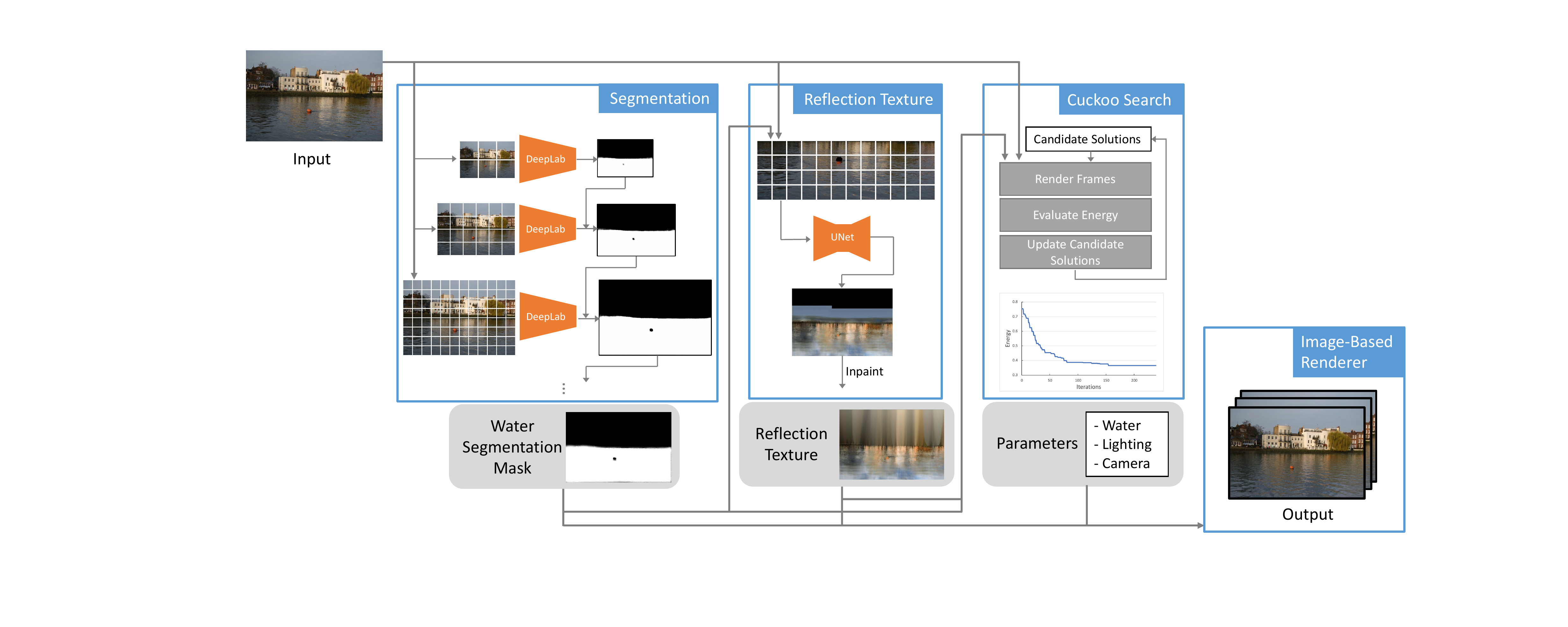}
    \caption{System overview. Given an input image, we first segment water, then predict reflection texture from the water segment, and estimate other parameters. With the reflection texture and parameters, we generate the water animation using an image-based renderer.}
    \label{fig:overview}
\end{figure*}

In this work, we are interested in automatically animating the water surface in a still photograph. In particular, we seek to generate and render the high-fidelity motion of large water bodies such as lakes and seas in a fully automatic manner. To avoid crossing into the ``uncanny valley'', previous methods either require significant user annotation to specify the water motion in the image~\cite{DBLP:journals/tog/ChuangGZCSS05,DBLP:journals/vc/OkabeDA18,le2022animating}, such as the dynamic region boundary and flow directions in the concurrent work of~\citet{le2022animating}, or assume the availability of video examples or multiple frames containing the target water motion~\cite{DBLP:conf/iccv/SunJF03,DBLP:journals/tvcg/LinWWKF07,DBLP:journals/cgf/OkabeAIS09,DBLP:journals/cgf/OkabeAO11,DBLP:journals/jzusc/GuiMYC12,DBLP:journals/cgf/PrashnaniNVS17} and use such inputs to drive image animation. The video-driven methods by~\citet{DBLP:conf/iccv/SunJF03,DBLP:journals/cgf/OkabeAIS09,DBLP:journals/jzusc/GuiMYC12,DBLP:journals/cgf/OkabeAO11} extract the motion field (\textit{e.g.} optical flow) from a driving video and transfer motion into the image, allowing the still water to come alive and mimic the motion in the driving video. To avoid the instability of optical flow,~\citet{DBLP:journals/cgf/PrashnaniNVS17} propose to use phase variations to model the motion. However, these methods cannot decouple appearance and dynamics in the target image, which limits their capacity in processing the water surface with distorted fluctuation and reflectance. The recent works~\cite{DBLP:conf/cvpr/TesfaldetBD18,DBLP:journals/tog/EndoKK19} decouple appearance and dynamics using two pre-trained convolutional neural networks (CNNs). \citet{DBLP:conf/cvpr/TesfaldetBD18} propose to transfer dynamics only from the driving video while preserving the appearance in the target image. The work by~\citet{DBLP:journals/tog/EndoKK19} predicts optical flows to tackle the single-image animation generation task. Since these methods use spatially-invariant statistics to represent the appearance, they are also limited to processing spatiotemporally homogeneous data. Recently, another impressive work~\cite{DBLP:conf/cvpr/HolynskiCSS21} proposes a motion representation based on Euler integration and synthesizes plausible motion for a given image by learning from a large-scale video dataset. These non-parametric methods synthesize dynamic water using a pure generation model, much like a black box, without building physically correct geometry and reflectance, thus limiting the resulting quality and resolution and also resulting in a lack of control regarding the appearance, diversity, and consistency of the animated water. 

Taking only a \textit{single} image of a target scene, we aim to animate and render the water regions. Our work lies in the confluence of traditional parametric models and learning-based non-parametric models of water generation in an attempt to combine the best of both worlds: the artifact-free rendering and flexible control of parametric models and the generalization ability of non-parametric models. We represent the water geometry and appearance using an empirical parametric model~\cite{Tessendorf2001}, and automatically estimate the model parameters from the input image using both optimization-based and learning-based methods. With the estimated parameters, we can simulate and render the water that is visually similar to the input photograph and directly use the parametric model for creation and editing of the water animation.

More specifically, the parameter estimation is the core of our method. It is a challenging ill-posed problem since we need a full set of parameters from a single input image, including numerical parameters of appearance (\textit{i.e.} color), dynamics (\textit{i.e.} wind speed, wind direction, and wave choppiness), cameras (\textit{i.e.} angle, height, and field of view) and environmental lighting (\textit{i.e.} spherical harmonics (SH)~\cite{DBLP:conf/siggraph/RamamoorthiH01a}), as well as the reflectance of the entire scene. To address these challenges, we first formulate the reflectance estimation as an image synthesis task of generating a reflection texture from the water image using deep neural networks. Similar methods have achieved great success in many related image-based prediction tasks from a single image, such as distortion correction~\cite{DBLP:conf/cvpr/LiZSL19,DBLP:journals/tog/LiZLS19}, reflection removal~\cite{DBLP:conf/cvpr/WenT0LHH19,DBLP:conf/cvpr/LiY0LH20}, and denoising~\cite{DBLP:conf/cvpr/GuoY0Z019,DBLP:conf/cvpr/WeiFY020}. Due to practical constraints, we synthesize water images with realistic reflection effects for the supervised learning. Next, to predict lighting and other water dynamics parameters, we propose an adaptation of the cuckoo search metaheuristic~\cite{DBLP:conf/nabic/YangD09}, due to its simplicity and flexibility in exploring different candidate solutions.

We combine these techniques of parameter estimation together into a novel system to animate the water surface in a still photograph, as shown in Fig.~\ref{fig:overview}. We first develop a progressive framework for water segmentation of a high-resolution image (Sec. \ref{sec:segmentation}). We then leverage a supervised learning method to predict reflectance information for
the water surface as needed by our real-time renderer (Sec. \ref{sec:reflection_texture}). Next, we use the cuckoo search meta-heuristic
to estimate water surface parameters with a parallelized energy evaluation scheme (Sec. \ref{sec:cukoo_search}). Equipped with the predicted reflectance and the estimated parameters, we customize a real-time renderer with an efficient image-based screen space local reflection method (Sec. \ref{sec:renderer}) to generate the final animation. We demonstrate the effectiveness of our method on a variety of scenes and interactive applications.

\begin{figure}[t]
\centering
\begin{minipage}{0.24\hsize}
    \centering
    \includegraphics[keepaspectratio, width=\hsize]{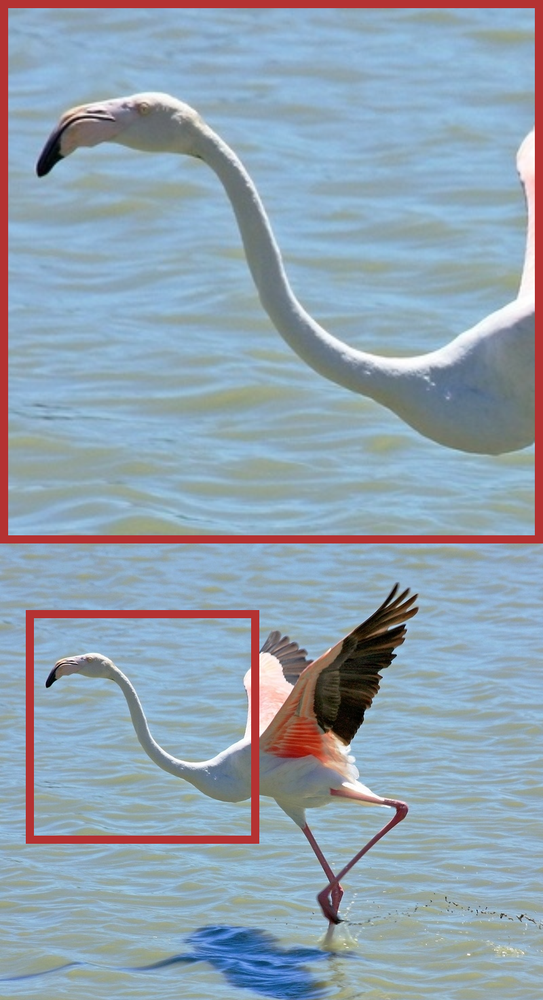}
    \subcaption{Input}
\end{minipage}
\begin{minipage}{0.24\hsize}
    \centering
    \includegraphics[keepaspectratio, width=\hsize]{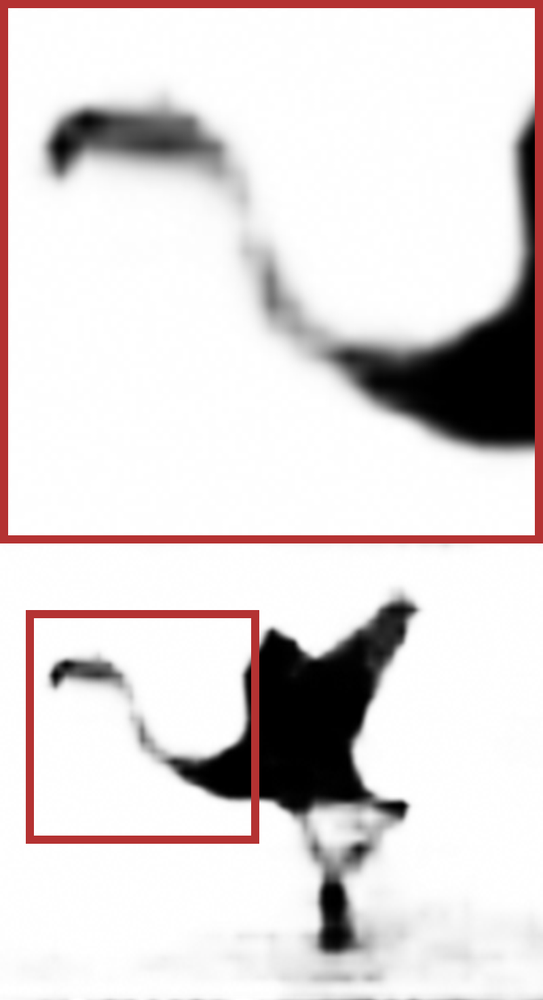}
    \subcaption{Level 1}
\end{minipage}
\begin{minipage}{0.24\hsize}
    \centering
    \includegraphics[keepaspectratio, width=\hsize]{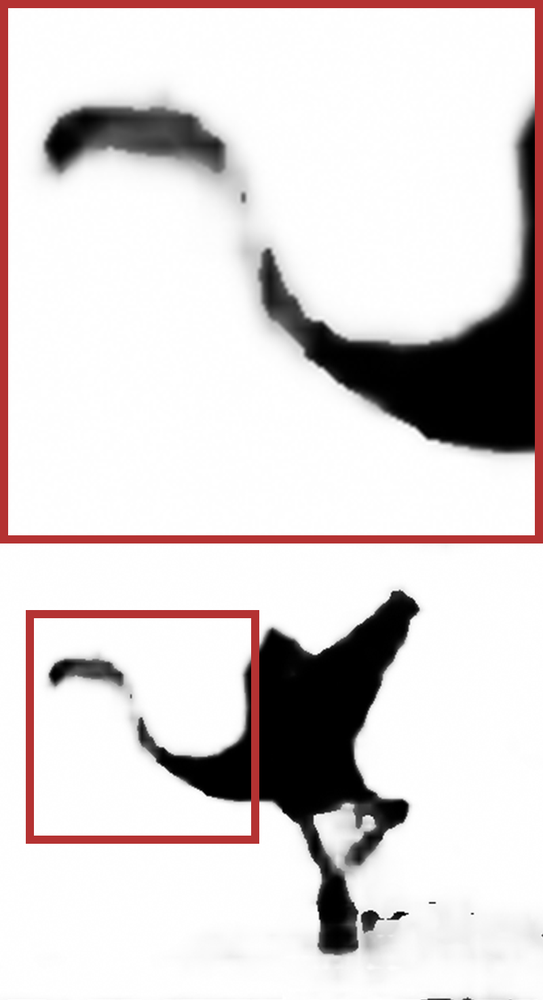}
    \subcaption{Level 2}
\end{minipage}
\begin{minipage}{0.24\hsize}
    \centering
    \includegraphics[keepaspectratio, width=\hsize]{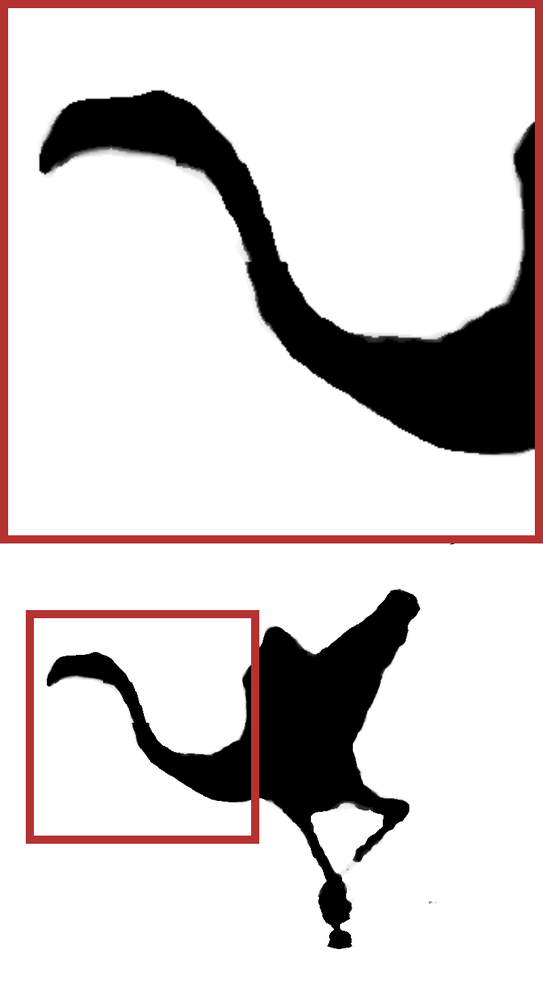}
    \subcaption{Level 3}
\end{minipage}
\caption{Intermediate results of the progressive patch-based segmentation framework, As the image level and resolution increase, fine-detailed segmentation edges are progressively detected. Input image: Andrea S/Flickr.} 
\label{fig:water_segment}
\end{figure}

\section{Water Segmentation} \label{sec:segmentation}
Semantic image segmentation is one of the fundamental topics in computer vision and recently has been solved effectively using deep convolutional neural networks deployed in a fully convolutional manner~\cite{DBLP:conf/cvpr/LongSD15,DBLP:journals/corr/ChenPSA17,DBLP:conf/cvpr/ZhaoSQWJ17,DBLP:journals/pami/ChenPKMY18,DBLP:conf/eccv/ChenZPSA18,DBLP:conf/eccv/ZhaoQSSJ18,DBLP:conf/cvpr/LiuCSAHY019}. Although the state-of-the-art systems propose to segment images at high resolution (usually $1K$ or $2K$) by either incorporating multi-resolution branches~\cite{DBLP:conf/eccv/ZhaoQSSJ18} or forming a hierarchical architecture search space in the network~\cite{DBLP:conf/cvpr/LiuCSAHY019}, their scalability is still limited by computing resources.

To segment a water image at higher resolution (\textit{e.g.} $4K$, $8K$ or higher in our application setting, we propose a progressive patch-based segmentation framework. We first build an image pyramid of multiple levels of detail by iteratively downscaling the input image to a resolution with the longest edge no large than 512 pixels at the lowest level. At each level, we dice the image into a set of patches ($512 \times 512$ pixels) in a 2D grid structure with $50\%$ overlap between adjacent patches. Then we start segmenting water on each patch from the lowest level using a pre-trained model of water segmentation, bilinearly upsampling the segmentation probability to the next level while iteratively updating the patches with pixel error larger than a threshold (we set it to $0.2$ in our implementation). For the pixels covered by multiple levels or multiple overlapping patches, we take the maximum probability as output. Our motivation is that low-level segmentation provides global information while the high-level segmentation ensures higher accuracy around boundaries. The iterative refinement repeats as the level increases until the image resolution reaches $4K$. Then for each patch, we upscale its predicted mask to the original resolution using the guided filter~\cite{DBLP:journals/corr/He015} to further refine the often over-smoothed segmentation result. Fig.~\ref{fig:water_segment} shows the segmentation result of each level and it can be seen that the segmentation boundary is clearer at higher levels. The water patch segmentation network is based on the advanced architecture of Deeplab~\cite{DBLP:journals/pami/ChenPKMY18}. For further details on the training, please refer to the supplemental material.

\section{Reflection texture generation} \label{sec:reflection_texture}
Next, we aim to generate a \textit{reflection texture}. This texture will later be used during shading of the the water in order to more accurately model reflections on the water surface. The rendering algorithm is described in Sec.~\ref{sec:renderer}.

When the water surface is as flat as a mirror, our approach seeks to generate a sharp reflection texture, preserving all high frequency details of the scene in the water reflection. In contrast, when the water surface is turbulent, the approach generates a blurry texture by resulting in a more realistic approximation to the complex interaction between light and turbulent wave dynamics.

The texture is built on image space and we seek to store the reflected color of the water surface as if the water was a completely flat mirror. The colors are then dilated to the regions beyond the water surface using an inpainting technique. Due to the high resolution, overlapping texture patches are first predicted by a patch-based learning network and then stitched together into a full texture map. 

\paragraph{Dataset}.
Our dataset consists of pairs of an image patch of water and its corresponding reflection texture patch. Since the ground truth of reflection texture is not available, we create a synthetic dataset. We use an image patch (as a reflection texture) and random parameters to render a water image patch with a size $224 \times 224$. Then, the rendered water image patch and its reflection texture patch are treated as a ground truth pair for training (see Fig. \ref{fig:ref_tex_dataset}). Further details on the dataset can be found in the supplemental material.

\begin{figure}[t]
\centering
\begin{minipage}{0.24\hsize}
    \centering
    \includegraphics[keepaspectratio, width=\hsize]{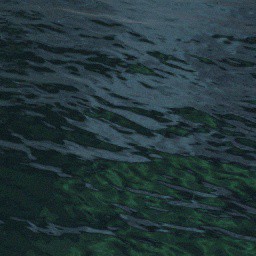}\\
    \includegraphics[keepaspectratio, width=\hsize]{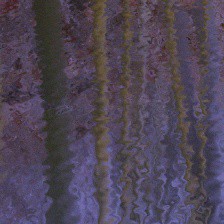}
\end{minipage}
\begin{minipage}{0.24\hsize}
    \centering
    \includegraphics[keepaspectratio, width=\hsize]{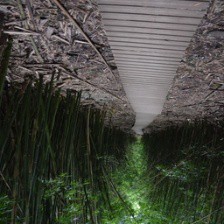}\\
    \includegraphics[keepaspectratio, width=\hsize]{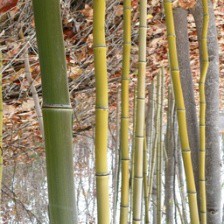}
\end{minipage}
\begin{minipage}{0.24\hsize}
    \centering
    \includegraphics[keepaspectratio, width=\hsize]{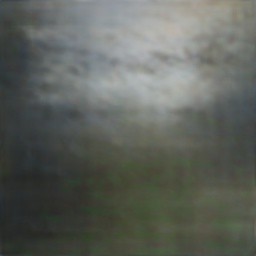}\\
    \includegraphics[keepaspectratio, width=\hsize]{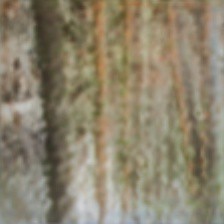}
\end{minipage}
\begin{minipage}{0.24\hsize}
    \centering
    \includegraphics[keepaspectratio, width=\hsize]{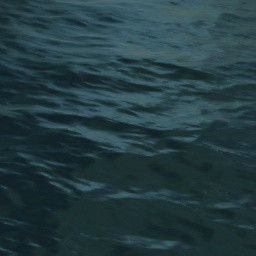}\\
    \includegraphics[keepaspectratio, width=\hsize]{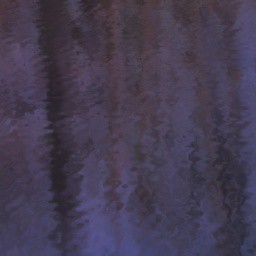}
\end{minipage}\\
\vspace{5pt}
\begin{minipage}{0.24\hsize}
\centering
\textbf{\small(a) Input}
\end{minipage}
\begin{minipage}{0.24\hsize}
\centering
\textbf{\small(b)~Ground~truth}
\end{minipage}
\begin{minipage}{0.24\hsize}
\centering
\textbf{\small(c) Predicted}
\end{minipage}
\begin{minipage}{0.24\hsize}
\centering
\textbf{\small(d) Rendered}
\end{minipage}
\caption{Images from the dataset along with their resulting predictions. The first row shows an example with turbulent water and a blurry reflection. The second row shows an example where the water is calm and the predicted texture is sharper. (d) shows the patches rendered using the predicted reflection textures (c). Blurry prediction suffices when the water is turbulent. Ground truth: Places~\cite{DBLP:conf/nips/ZhouLXTO14}.} 
\label{fig:ref_tex_dataset}
\end{figure}

\paragraph{Network architecture}. Our network learns a mapping from the rendered image patch to its reflection texture patch. Despite the reduced rendering model and difference between the synthetic images and real images, this method allows us to generate a reflection texture from any image with a water surface. The network architecture is based on UNet~\cite{DBLP:conf/miccai/RonnebergerFB15} with residual blocks~\cite{DBLP:conf/cvpr/HeZRS16} and multi-level skip connections between the encoder and decoder that can preserve sharp edges in reflection textures when desired. Note that the blurred reflection texture that is generated for turbulent water does not often degrade the quality of rendering when used, since they will be applied using turbulent water parameters. Refer to the supplemental material for further details on the network architecture.

\paragraph{Loss evaluation}.
During training, we distinguish water and non-water regions using a random mask image $m$ drawn from a set of mask images and their inverses generated using the ground truth annotation of the COCO dataset~\cite{DBLP:conf/eccv/LinMBHPRDZ14}. We apply the mask to the input patch (i.e. multiply the mask pixel value to the input image pixel at each pixel location) and feed it to the network to get a prediction patch. Notice that the range of mask value is $[0, 1]$.
This is to simulate real input patches, which may have both water and non-water regions. Then, the loss is defined as\begin{equation}
    L(x, y) = \frac{1}{N}\sum_{i}^N (m_i + \lambda (1-m_i))|x_i-y_i|,
\end{equation}
\noindent where $N$ is the number of pixel locations, $x_i, y_i$ represent the pixel color of the two patches $x$ and $y$ at the pixel location $i$ respectively, and $m_i$ is the mask value at that pixel location. We let $\lambda=0.1$ to impose a small loss on pixels that are outside of water regions. We do not expect the network to fill non-water regions with accurate reflection colors; we need the network to fill such regions with colors that transition smoothly from the neighbouring water regions.

\paragraph{Stitching and inpainting.}
During testing, we partition the input image and the mask into patches with $80\%$ overlap in each dimension and feed each image patch into our network to get its corresponding reflection texture patch. These reflection texture patches are stitched together using weights following a Gaussian kernel within the overlap region to ensure a smooth transition. After stitching, we inpaint the non-water regions of the full reflection images using the method of \cite{Telea2004}. This is needed since  we may occasionally need to fetch the reflection color slightly outside of the water region in cases where the water surface is turbulent.

\begin{figure}[t]
    \centering
    \begin{minipage}{0.32\hsize}
        \centering
        \includegraphics[keepaspectratio, width=\hsize]{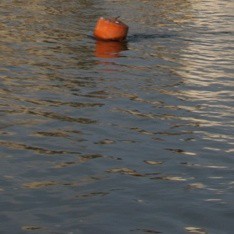}
        \subcaption{Input}
    \end{minipage}
    \begin{minipage}{0.32\hsize}
        \centering
        \includegraphics[keepaspectratio, width=\hsize]{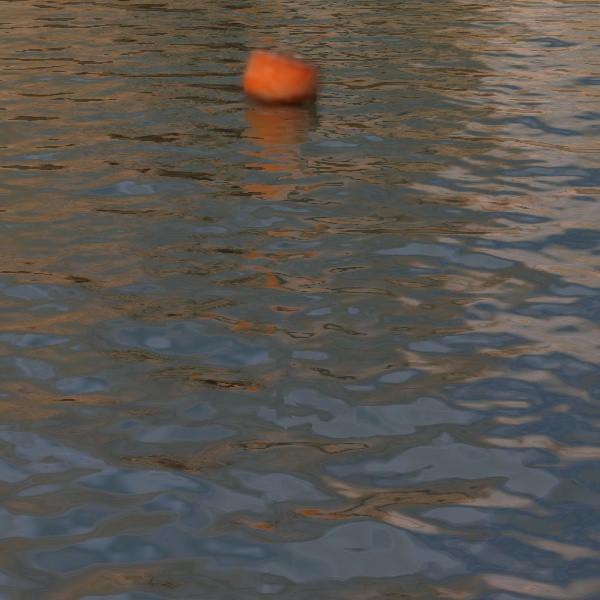}
        \subcaption{Our method}
    \end{minipage}
    \begin{minipage}{0.32\hsize}
        \centering
        \includegraphics[keepaspectratio, width=\hsize]{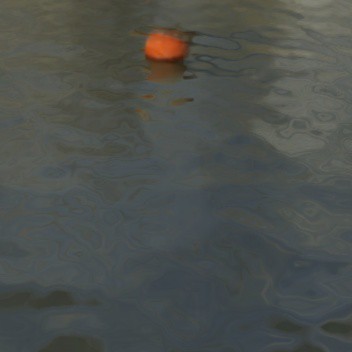}
        \subcaption{Neural network}
    \end{minipage}
    \caption{Comparison between our method and a neural network approach. The result of the neural network method in this example lacks the high frequency waves present in the input image. Furthermore, the overall color is shifted towards green.}
    \label{fig:nn_fail}
\end{figure}

\section{Parameter Estimation by Cuckoo Search} \label{sec:cukoo_search}
In addition to the segmentation mask and the reflection texture presented in the previous sections, the rendering algorithm takes a 21-dimensional vector of wave, wind, camera, and lighting parameters. The supplemental material contains a table with all of the parameters and their respective ranges.

We initially tried to design a neural network similar to the reflection texture generation network that performed parameter inference by creating a synthetic dataset that included ground truth parameters. However, the predicted parameters did not produce plausible results when used for rendering, as shown in Fig.~\ref{fig:nn_fail}. The problem is ill-posed as different sets of parameters may give a similar output (e.g.\ high wind speed with a high camera position vs. low wind speed with a low camera position). We believe that this ambiguity makes it challenging for the network to learn the parameters accurately. We investigated different techniques to explore the space of possible parameter solutions including traditional optimization methods and learning-based approaches. Finally, we have found an adaption of the cuckoo search metaheuristic \cite{DBLP:conf/nabic/YangD09} to be a suitable choice for our ill-conditioned and non-differentiable problem. This is because the cuckoo search does not assume any specific characteristics of the optimization problem such as convexity and does not require a gradient of the solution.

In our application, we use cuckoo search with an energy function based on a combination of the DISTS similarity metric \cite{DBLP:journals/corr/abs-2004-07728} and an HSV color histogram metric. For each candidate parameter set, we render an image and calculate a distance between rendered image $y$ and the original input image $x$. We find a parameter set that minimizes this distance.

\begin{figure}[t]
    \centering
    \begin{minipage}{0.32\hsize}
        \centering
        \includegraphics[keepaspectratio, width=\hsize]{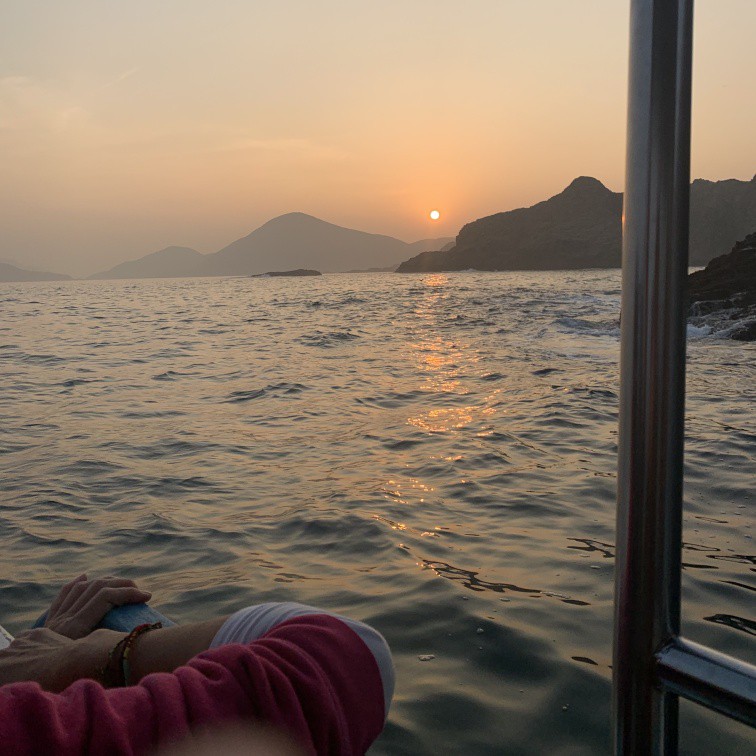}
        \subcaption{Input}
    \end{minipage}
    \begin{minipage}{0.32\hsize}
        \centering
        \includegraphics[keepaspectratio, width=\hsize]{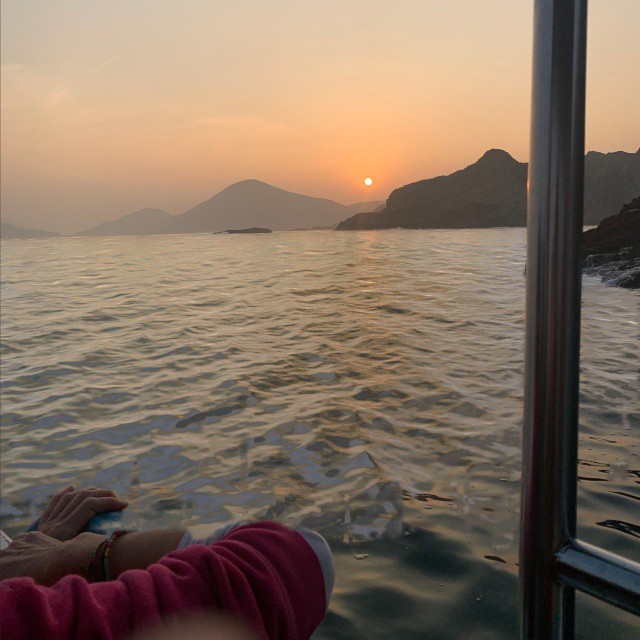}
        \subcaption{Our method}
    \end{minipage}
    \begin{minipage}{0.32\hsize}
        \centering
        \includegraphics[keepaspectratio, width=\hsize]{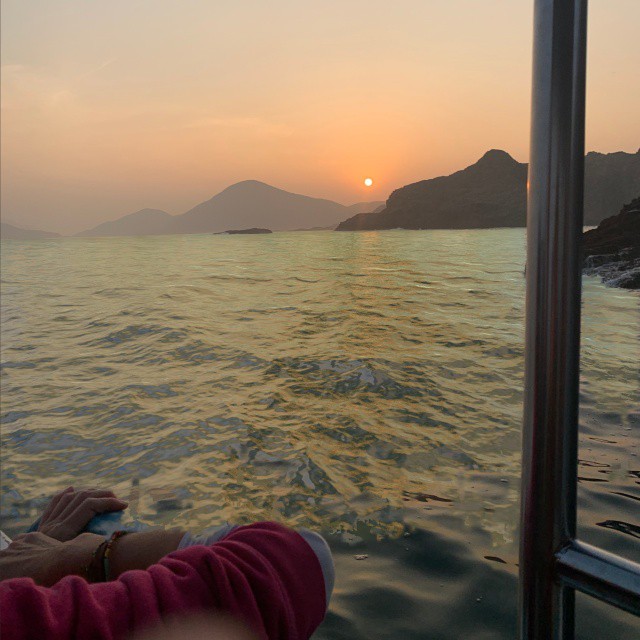}
        \subcaption{w/o color metric}
    \end{minipage}
    \caption{Comparison between our method with and without the color similarity metric.}
    \label{fig:hsv_match}
\end{figure}

\subsection{Energy Function}\label{sec:energy}
As mentioned above, our energy function considers both the DISTS metric and a color histogram metric:
\begin{equation}
E_T + \lambda E_C\ ,
\end{equation}
where $E_T$ is DISTS energy, $E_C$ is the color dissimilarity energy, and $\lambda$ regulates the tradeoff ($\lambda = 1.0$ in our implementation).

\paragraph{Texture similarity using DISTS index.}
Since the water surface is dynamically changing, we need a distance metric that is not sensitive to local variations yet globally consistent with human perceptual scores. We have found that the DISTS index is suitable for this task and apply it directly in our approach:
\begin{equation}
    E_T = d(x, y)\ ,
\end{equation}
\noindent where $d(x, y)$ is the DISTS index measuring the dissimilarity between images $x$ and $y$.

\paragraph{Color similarity using HSV color histogram.}
While the DISTS metric achieves good results in measuring similarity in the overall structure of the image content, it has not been as effective in ensuring color similarity. Thus we added a color histogram distance metric, which measures similarity in HSV color space. More specifically, the images are converted to HSV color space, and then for each image, all pixels are classified into one of $24 \times 8 \times 8$ partitions based on hue (24 classes), saturation (8), and value (8). Note that we allocate more partition resolution to the hue. The distance between two sets of bins is then measured using the Hellinger distance, which is an effective technique to measure the amount of overlap between two distributions:
\begin{equation}
    E_C = H(x, y) = \sqrt{1-\frac{1}{\sqrt{\sum_{i=1}^n x_i \sum_{i=1}^n y_i}}\sum_{i=1}^n \sqrt{x_i y_i}}\ ,
\end{equation}
\noindent where $n$ is the number of partitions, and $x_i, y_i$ represent the number of pixels that fall into partition $i$ from images $x$ and $y$, respectively. In Fig.~\ref{fig:hsv_match}, we compare our final method with and without considering the HSV color similarity metric. Note that the addition of the color similarity results in renderings that are much more consistent with the colors of the input image.

\paragraph{Evaluation details.} 
The input to our evaluation is a set of two images: the original input image and an image rendered with a candidate parameter set. First, we crop the bounding box for all water regions for both the original image and reflection texture and resize them to 256x256. Then, using the resized reflection texture and the candidate parameter set, an image of the same resolution is rendered. The rendered image and the resized original image are passed to the two metrics for evaluation. 

\subsection{Algorithm}

A simple cuckoo search maintains a set of $n$ nests (25 in our implementation), each with a candidate solution, or egg (\textit{i.e.}, values for each of the parameters). In each iteration, a new solution, or cuckoo egg, is generated for each nest via a L\'{e}vy flight, a random walk in the parameter space in which the step size follows the L\'{e}vy distribution. The cuckoo egg replaces an egg in another nest if it improves upon the latter. Next, for each nest, a new cuckoo egg is generated by mutation and replaces the egg in the nest if it is of improved quality. At the end of each iteration, the algorithm keeps the best nests and drops a small fraction $k$ of the nests (5 in our implementation) replacing them with new random eggs. The supplemental material describes a parallelized version of the algorithm for improved efficiency and provides the pseudocode for both.

\paragraph{Termination condition.}
The algorithm terminates when the optimization no longer yields significant improvements to the best egg.  Let the energy of the best egg at the end of iteration $k$ be $E(k)$. To reduce the influence of random oscillations, we first apply a smoothing filter to $E(\cdot)$, yielding the smoothed energy $E'(\cdot)$:
\begin{equation}
    E'(k) = \sum_{i=k-s+1}^{k} \{i-(k-s)\}E(i)\ ,
\end{equation}
where $s$ is the filter size. When $\{E'(k) - E'(k-1)\}/E'(k) < \epsilon$, the algorithm terminates. We use $s = 100$ for smoothing and $\epsilon = 0.0001$ as a  conservative enough threshold.

\section{Renderer} \label{sec:renderer}
Our algorithm renders realistic deep water bodies with local reflections, taking as input the the segmentation mask and reflection texture as well as the wave, wind, camera, and lighting parameters mentioned earlier and listed in the supplemental material.

Realistic simulation of water has been researched extensively in the graphics community. The methods can be classified into two main categories: physically-based models and empirical methods \cite{Darles2011}. Physically-based methods are suitable for shallow water simulation and are capable of simulating full three dimensional behavior of water \cite{Bridson2007}. Empirical methods generate 2D displacement maps of water surface of deep water scenes. Empirical methods can be further categorized into spatial-domain methods, spectral-domain methods, and hybrid methods \cite{Darles2011}. In our work, we employ the spectral domain approach presented in \citet{Tessendorf2001} to generate a displacement map of water surface using the wave and wind parameters. This is a commonly used approach in film production and real-time applications.

Our approach considers the fact that the surface of the water may extend infinitely and also allows the user to interactively zoom. Thus the na\"{i}ve approach of constructing a large and dense mesh is not practical. We instead adopt the projected grid LOD method from \citet{Johanson2004}. More specifically, in the vertex shader, we generate a screen space mesh. We project each vertex to the water surface plane and apply the displacement based on the computed displacement map before projecting back to screen space. This results in a nearly uniformly and dense distribution of mesh vertices in screen space.

\begin{figure}[t]
\begin{center}
    \centering
    \begin{minipage}{0.49\hsize}
        \centering
        \includegraphics[trim=0 120 150 0, clip,width=\linewidth]{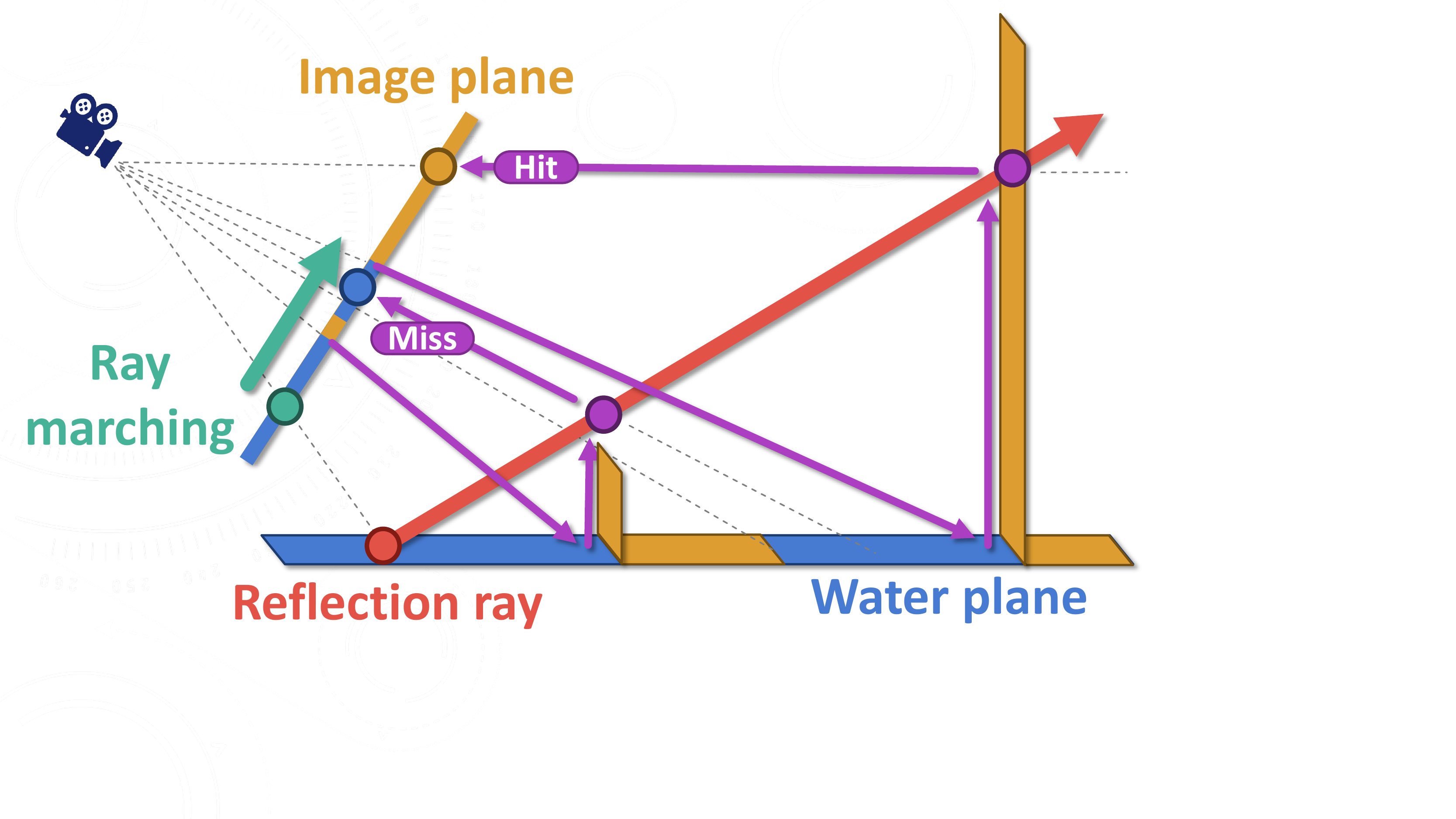}
        \subcaption{ray marching}
    \end{minipage}
    \begin{minipage}{0.49\hsize}
        \centering
        \includegraphics[trim=0 120 150 0, clip,width=\linewidth]{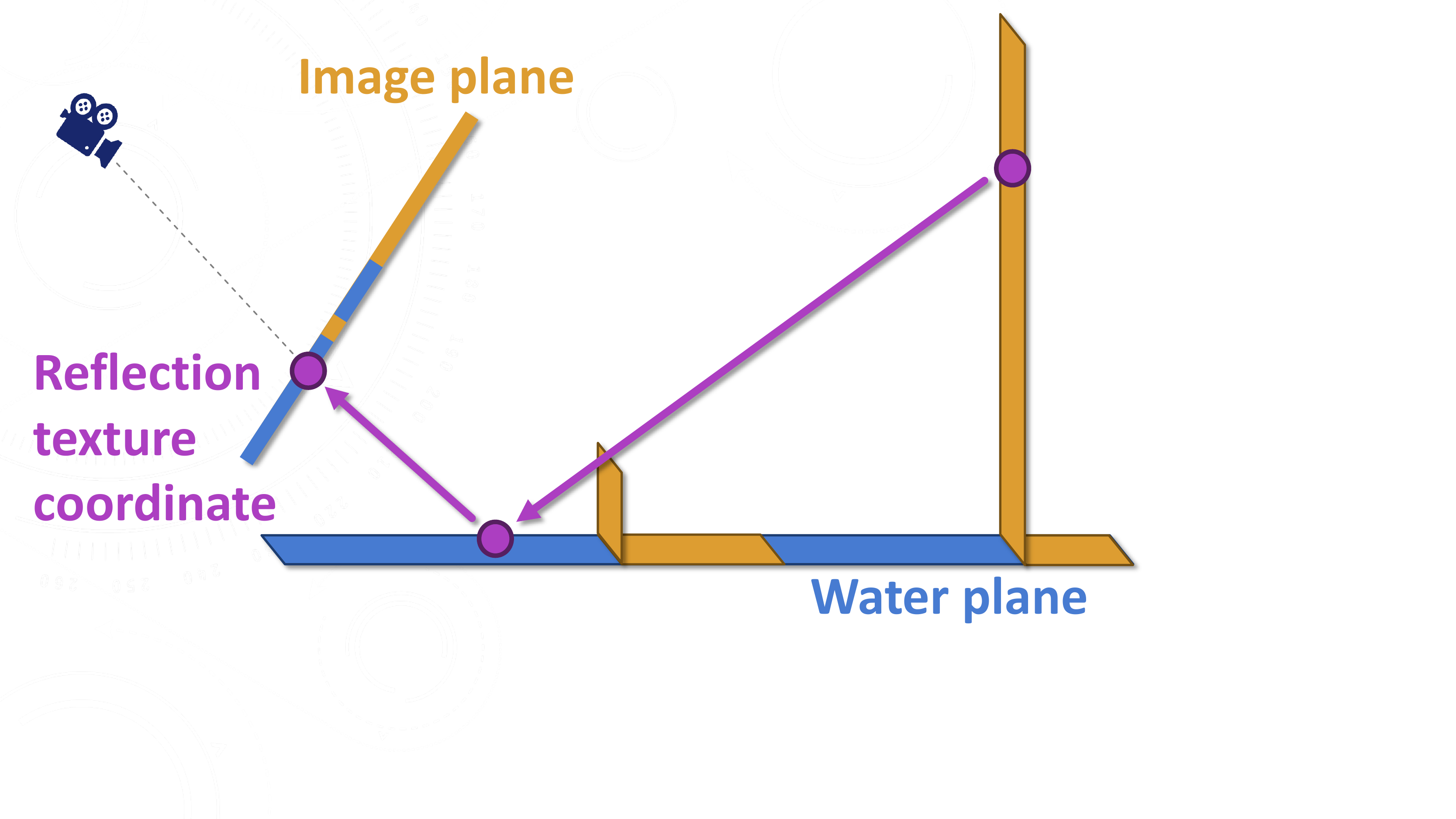}
        \subcaption{texture sampling}
    \end{minipage}
    \caption{The image-based reflection method. The objects are mapped onto the water plane to determine the reflection source point in 3D. We perform ray-marching in image space with additional checks in 3D space to consider the height of wall proxy objects. Then, we fetch the reflection color embedded in the water surface from the reflection texture.}
    \label{fig:wall}
\end{center}
\end{figure}

\paragraph{Image-based reflection.}
Water surface reflection is an essential component in making the water surface appear realistic. \citet{Tessendorf2001} provides a good summary on surface wave optics. While common implementations to achieve this goal use environment mapping, it is challenging to estimate an environment map from one static image and apply it to the entire scene. 
Furthermore, environment maps have a limitation of not being able to simulate local reflections. We take a similar approach to the screen space reflection method of \citet{McGuire2014}. This is a practical method used for real-time applications which computes reflection given the normal and depth information from the scene. In our application, we do not have this information directly, but we can approximate it given the camera pose and water mask. Also, rather than retrieving the reflection color from the surface of objects, we make use of the reflection component embedded within the water regions stored in the precomputed reflection texture to allow reflection estimations even when objects are absent in the input image. %

We first place {\em vertical curved walls} along water boundaries in world coordinates as proxies for 3D objects. The exact positions of the walls are automatically determined by projecting the water mask on the image plane onto the ground-level plane using the estimated camera parameters (see Fig.~\ref{fig:wall} for a diagram with a simple example). For each pixel, we compute the collision between the reflection ray from water facet and a vertical wall in world space. We compute the texel coordinate of the reflection texture which contains the corresponding reflection color. If the reflection ray completely misses the vertical walls, we retrieve a color from the closest point in the reflection texture. This approximation does not affect the quality of the final image significantly. This follows from the fact that there is very little contribution from the reflection when the incident angle is small. Since that is usually the case when the reflection vector does not hit any of the curved planes, the color of the water facet is dominated by the refraction term.

More specifically, these are the steps to compute the reflection color as implemented in our fragment shader. Further optimizations to improve running time are discussed in the supplemental material.
\begin{enumerate}[leftmargin=*]
\item {\em Calculate the reflection vector.} Given the view and normal vectors, we compute the reflection vector in world space and project it onto screen space.
\item {\em Compute wall collision points.} We then apply ray marching in screen space to find the collision points with water boundaries. Note that there can be multiple collisions per ray as shown in Fig.~\ref{fig:wall}a. We project the collision points back to world coordinates such that the points now reside on the water boundaries in world coordinates (forward purple arrows). We then calculate the collision points of the reflection vector and the object using the direction of the original reflection vector.
\item {\em Determine first valid collision point.} If a collision point lies on the vertical wall, the collision point is a valid reflection color source point. To determine whether the wall is high enough to result in a collision, we use an approximation that projects the collision point back to screen space and checks the mask value of that point to ensure it is not on the water surface. Then, we choose the valid collision point that is closest to the ray marching starting point.
\item {\em Calculate the reflection texture coordinates.}
We then calculate the point on the water surface plane that contains the reflection color under a flat mirror assumption (Fig.~\ref{fig:wall}b). Finally, we sample the reflection texture at that position to retrieve the color.
\end{enumerate}

\section{Experiments}\label{sec:results}

We test our system on a dataset consisting of 67 images (32 from Places~\cite{DBLP:conf/nips/ZhouLXTO14} and 35 in-the-wild images) with a variety of water scenes including oceans, rivers, ponds, and lakes. The image resolutions ranged from 748$\times$421 to 4032$\times$3024. All experiments are performed on a PC with an 8 core Ryzen 2700X CPU, 16GB of RAM, and an NVIDIA GeForce RTX 2070 GPU. Please refer to the accompanying video for animated renderings and additional results.

\paragraph{Runtime.} The runtime efficiency of our approach is proportional to the number of pixels in the water regions of the input image. For a $4K$ image (4032$\times$3024) with the water occupying approximately one-third of the image, performing the water segmentation takes approximately 7 seconds, predicting the reflection texture takes 9 seconds, and estimating the parameters using the cuckoo search metaheuristic takes 4.5 minutes to evaluate about 19000 different candidate solutions. About 60\% of execution time for optimization is spent on rendering, 30\% on the DISTS metric evaluation, and 10\% on color metric evaluation. The execution time for the optimization is indeed independent of input image size, as we evaluate the energy using a fixed-size image patch as described in Sec.~\ref{sec:energy}. A theoretical convergence analysis of the cuckoo search is out of scope for this paper and is still an open problem except for a simplified version of the algorithm~\cite{He2018}. We implement a real-time renderer using WebGL for users to view and interact with the resulting animation. Our optimized renderer reaches 50-60fps at $4K$ resolution.

\begin{figure}[t]
\includegraphics[width=\linewidth]{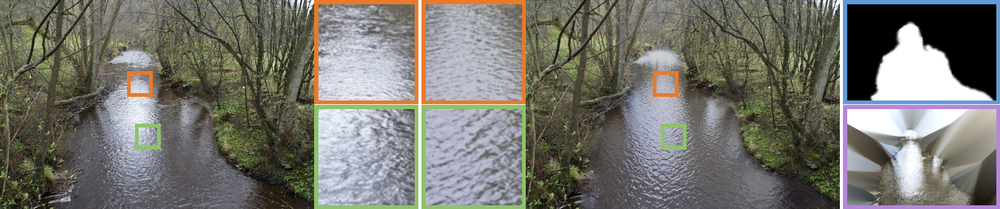}
\includegraphics[width=\linewidth]{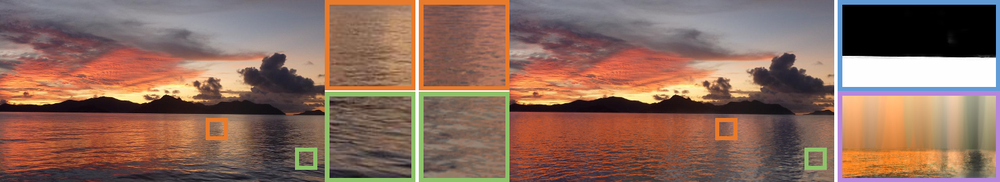}
\includegraphics[width=\linewidth]{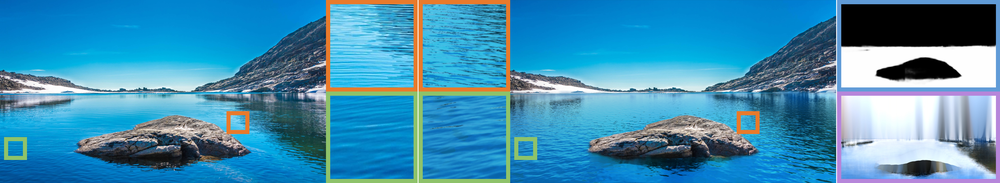}
\includegraphics[width=\linewidth]{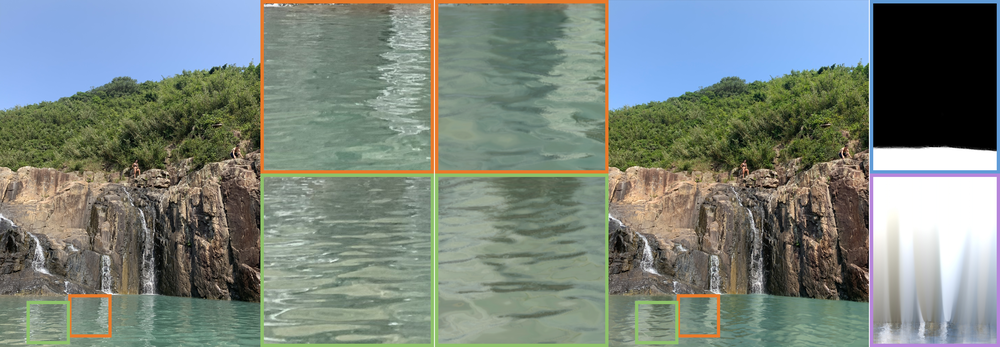}
\includegraphics[width=\linewidth]{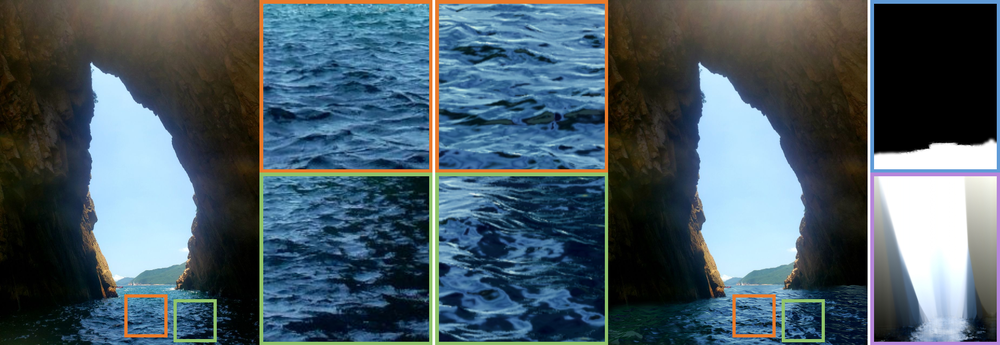}
\caption{Rendered results and closeups for different input scenes, along with the predicted segmentation masks and reflection textures. The input images are shown in the first column. Input images: Edward Nicholl/Flickr (first), Foliez/Pixabay (second), and Unknown/PxHere (third).}
\label{fig:results}
\end{figure}

\paragraph{Results.} We demonstrate the final results on a variety of input images generated by our method in Figs.~\ref{fig:teaser} and~\ref{fig:results}. We also visualize some close-up details, the predicted segmentation mask, and the reflection texture. Note the high accuracy of the mask, particularly for large water bodies of varying shapes. Our reflection texture network is able to generate a high-quality reflection texture for images with both calm and turbulent waters. Naturally, for calm waters, the reconstructed reflection is much sharper than for turbulent waters as shown in the third case of Fig.~\ref{fig:results}. Recall that in most cases, to render the turbulent water, a smooth reflection texture is sufficient since reflection details are not discernible in a turbulent setting.

\paragraph{User study.} In addition to the visual demonstration, we conduct a user study to evaluate the realism of our composited results. We randomly select 30 images from the testing dataset and create a poll of 30 ``real or rendered'' questions for each user. For every question, the user is randomly shown either the real input or the rendered result and then asked whether the image looks realistic or not. We retrieve 2,160 answers from 72 users and compute the ``real rate'', which is the percentage of images that the users tag as real. The real rate for real images is 81.67\% while the real rate for rendered images is 65.46\%, indicating that most of our generated results are thought to be as realistic as real images. As shown in this user study, although our method can hardly ensure pixel-level accuracy with respect to the ground truth image, it is able to generate photorealistic results with high fidelity at a low rendering cost for real-time applications. 

\begin{figure}[t]
    \centering
    \begin{minipage}{0.323\hsize}
        \centering
        \includegraphics[keepaspectratio, width=\hsize]{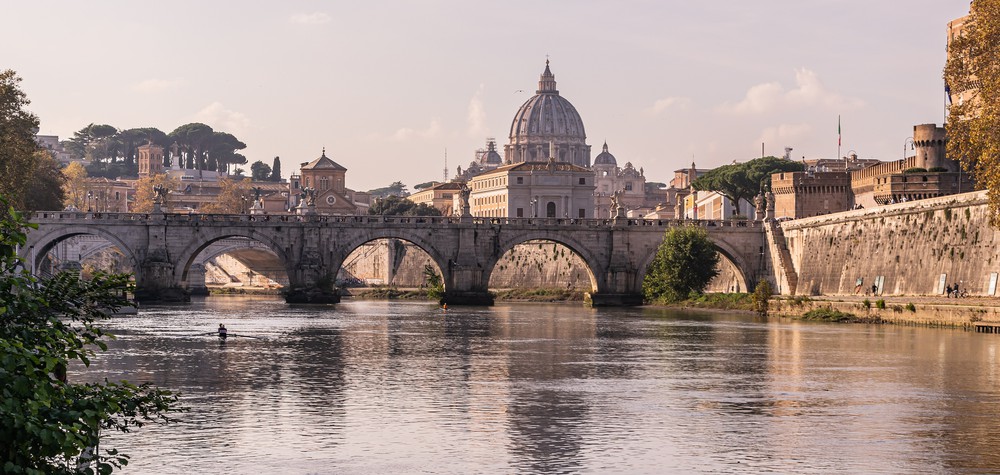}
    \end{minipage}
    \begin{minipage}{0.323\hsize}
        \centering
        \includegraphics[keepaspectratio, width=\hsize]{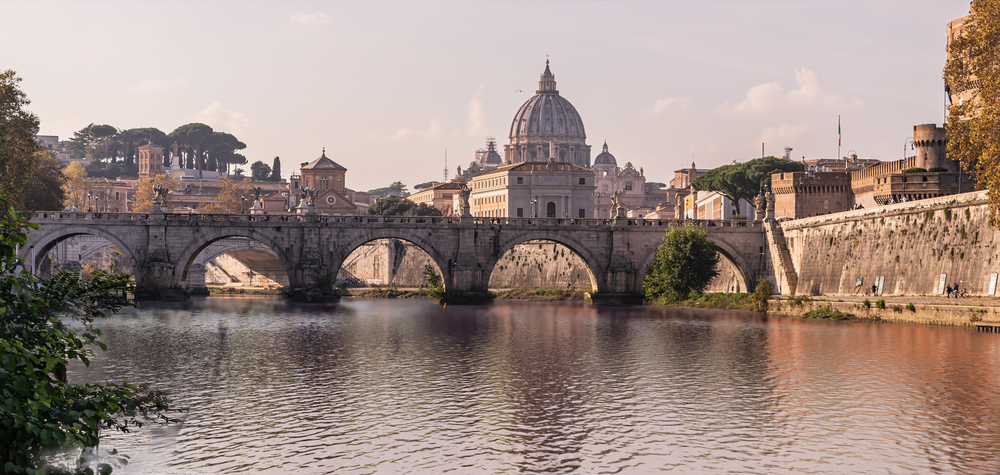}
    \end{minipage}
    \begin{minipage}{0.323\hsize}
        \centering
        \includegraphics[keepaspectratio, width=\hsize]{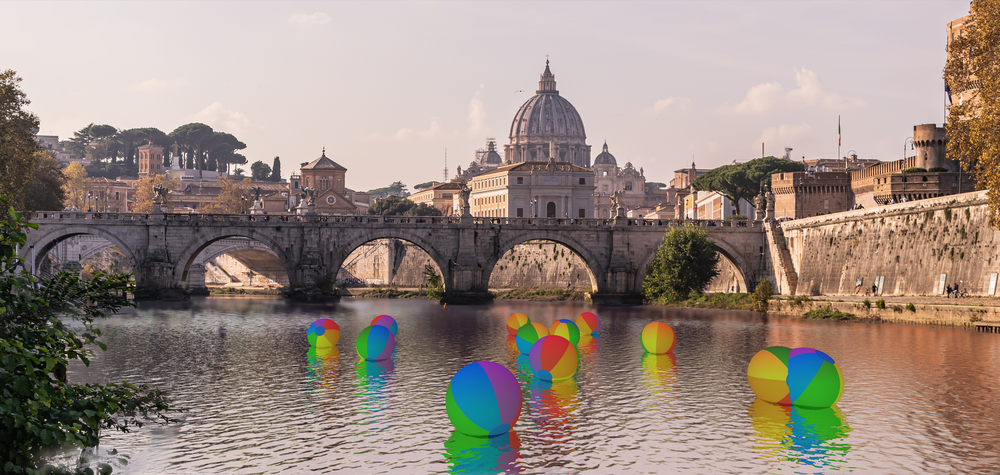}
    \end{minipage}\\
    \begin{tabular}{C{0.3\linewidth}C{0.3\linewidth}C{0.3\linewidth}}
    \footnotesize{\textsf{Input}} &\footnotesize{\textsf{Rendered}} &\footnotesize\textsf{{Edited}}\\
    \end{tabular}\\
    \footnotesize{\textsf{(a) Synthetic object insertion}}\vspace{5pt}\\
    \begin{minipage}{0.24\hsize}
        \centering
        \includegraphics[keepaspectratio, width=\hsize]{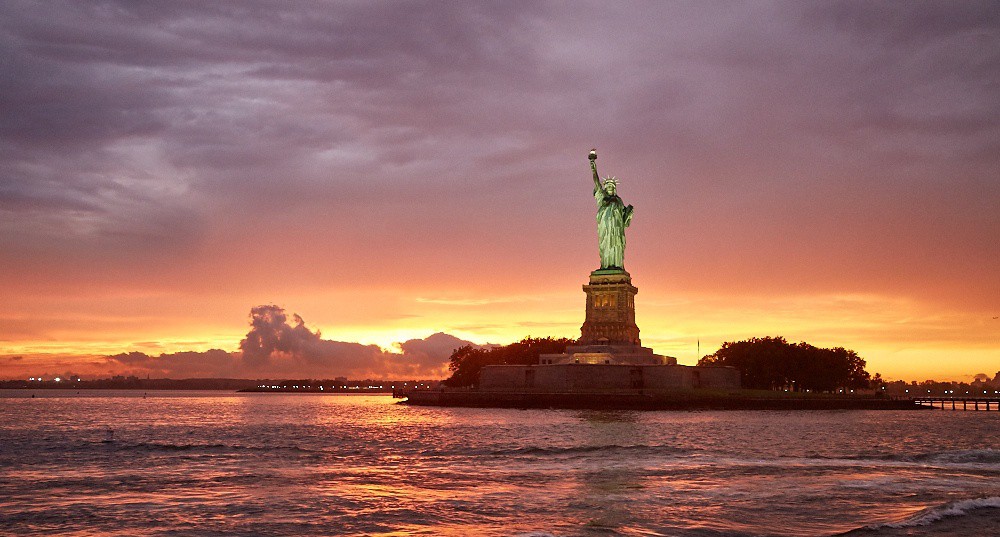}
    \end{minipage}
    \begin{minipage}{0.24\hsize}
        \centering
        \includegraphics[keepaspectratio, width=\hsize]{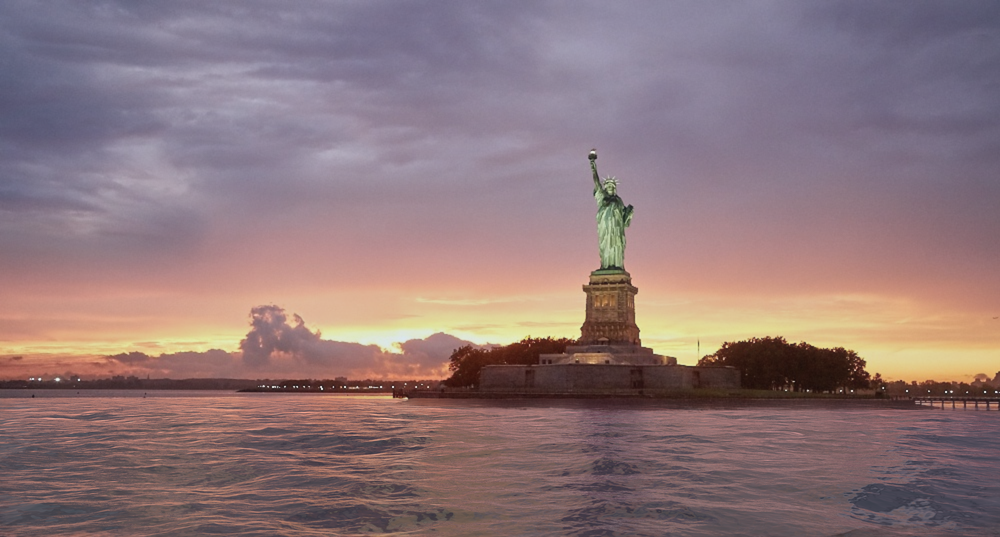}
    \end{minipage}
    \begin{minipage}{0.24\hsize}
        \centering
        \includegraphics[keepaspectratio, width=\hsize]{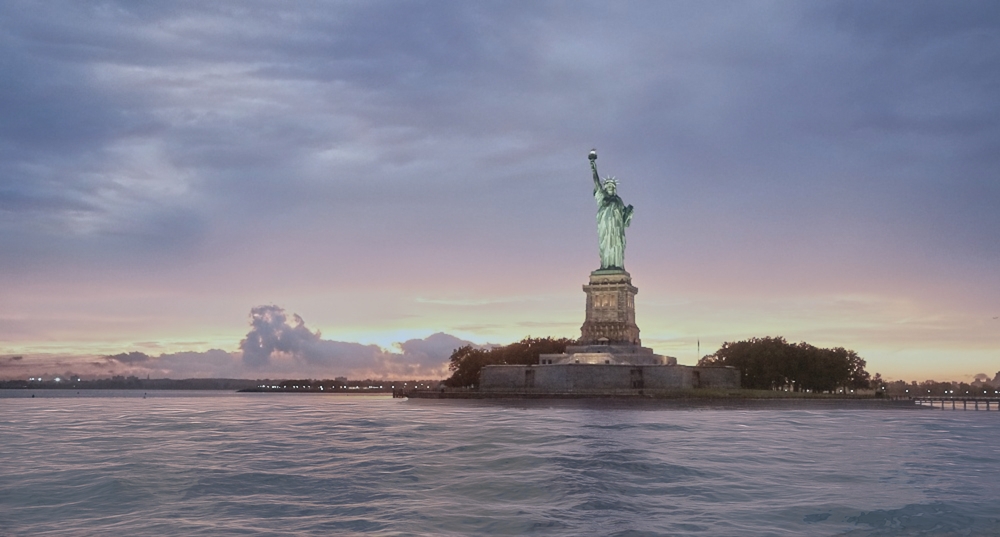}
    \end{minipage}
    \begin{minipage}{0.24\hsize}
        \centering
        \includegraphics[keepaspectratio, width=\hsize]{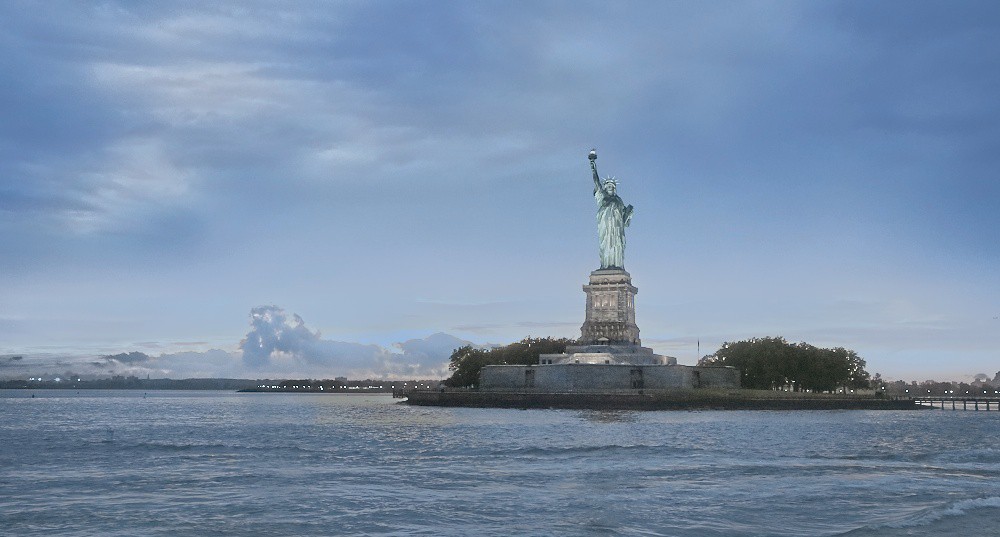}
    \end{minipage}\\
    
    \begin{minipage}{0.24\hsize}
        \centering
        \includegraphics[keepaspectratio, width=\hsize]{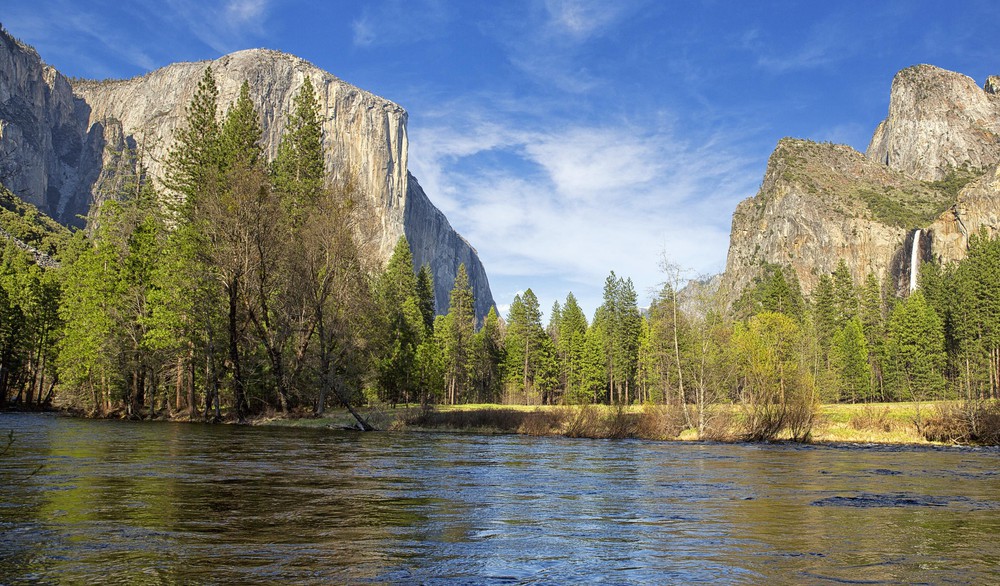}
    \end{minipage}
    \begin{minipage}{0.24\hsize}
        \centering
        \includegraphics[keepaspectratio, width=\hsize]{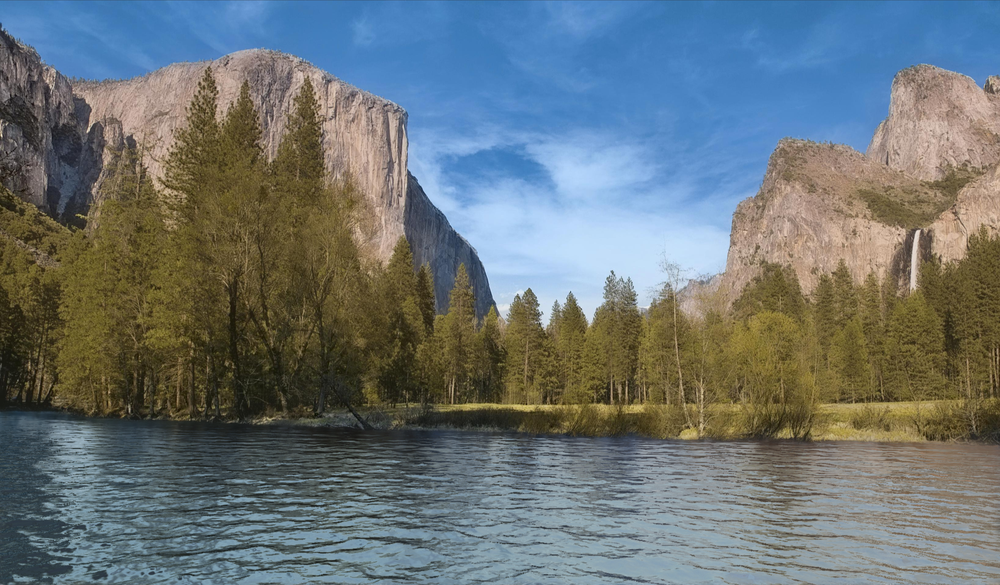}
    \end{minipage}
    \begin{minipage}{0.24\hsize}
        \centering
        \includegraphics[keepaspectratio, width=\hsize]{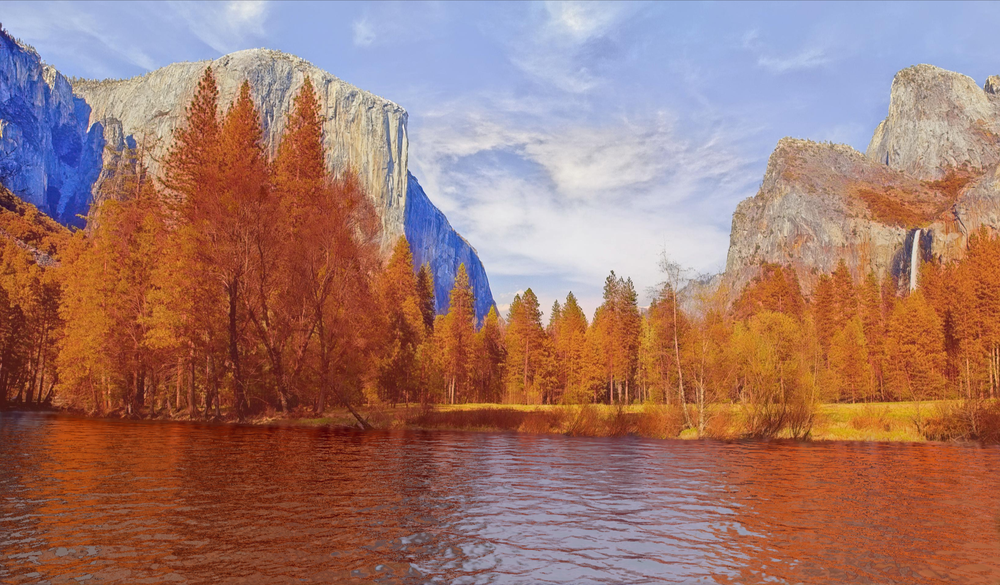}
    \end{minipage}
    \begin{minipage}{0.24\hsize}
        \centering
        \includegraphics[keepaspectratio, width=\hsize]{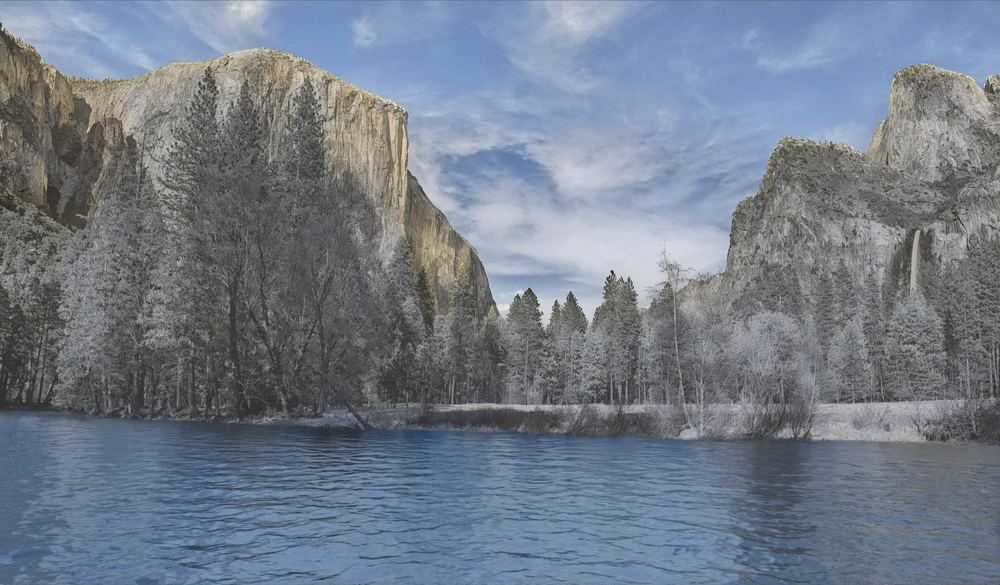}
    \end{minipage}
    
    \begin{tabular}{C{0.2\linewidth}C{0.215\linewidth}C{0.21\linewidth}C{0.21\linewidth}}
    \footnotesize{\textsf{Input}}&\footnotesize{\textsf{Edited 1}}&\footnotesize{\textsf{Edited 2}}&\footnotesize{\textsf{Edited 3}}\\
    \end{tabular}\\
    \footnotesize{(b) Reflection-aware color transfer}
    \caption{Applications. In (a), our method allows to place synthetic objects on the water surface and generates realistic reflections. In (b), given an input image, we transfer its color to simulate different environments and predict corresponding reflection textures. By interpolating between the predicted parameters, the edited results vary smoothly. Input Images: (a) Krzysztof Golik/Wikimedia (b) Confaulk/Wikimedia (top) and Rixie/Adobe Stock (bottom).}
    \label{fig:app}
\end{figure}

\paragraph{Applications.} Our method generates 3D representations of the water surface. It not only allows direct editing of parameters to control wave, wind, and lighting but also enables interesting applications such as insertion of synthetic objects and reflection-aware color transfer as shown in Fig.~\ref{fig:app}. To render the synthetic objects into the water (Fig.~\ref{fig:app}a), we first use the estimated camera pose and wave parameters to compute reflection and refraction vectors in world coordinates. Then, we compute intersections between the rays and the synthetic objects. The objects are shaded based on their normals and our estimated SH coefficients. In the reflection-aware color transfer (Fig.~\ref{fig:app}b), we can simulate consistent water animation as the color of the surrounding environment changes producing interesting time-lapse videos as shown in the supplemental video. Given an input image, we first apply a color transfer method~\cite{DBLP:journals/tog/HeLCYS19} to simulate its appearance at different times of the day or seasons. Taking the original image and one or more of its recolored results as input, we separately estimate the parameters and textures for each input image. We then use the median values among all the estimates for the wave dynamics and camera pose parameters, which circumvents artifacts due to any minor differences in the estimated parameters among all input images. Finally, we linearly interpolate the reflection textures and lighting parameters to generate a water animation with smoothly varying lighting conditions.

\section{Limitations and Summary}
\begin{figure}[t]
    \centering
    \begin{minipage}{0.16\hsize}
        \centering
        \includegraphics[keepaspectratio, width=\hsize]{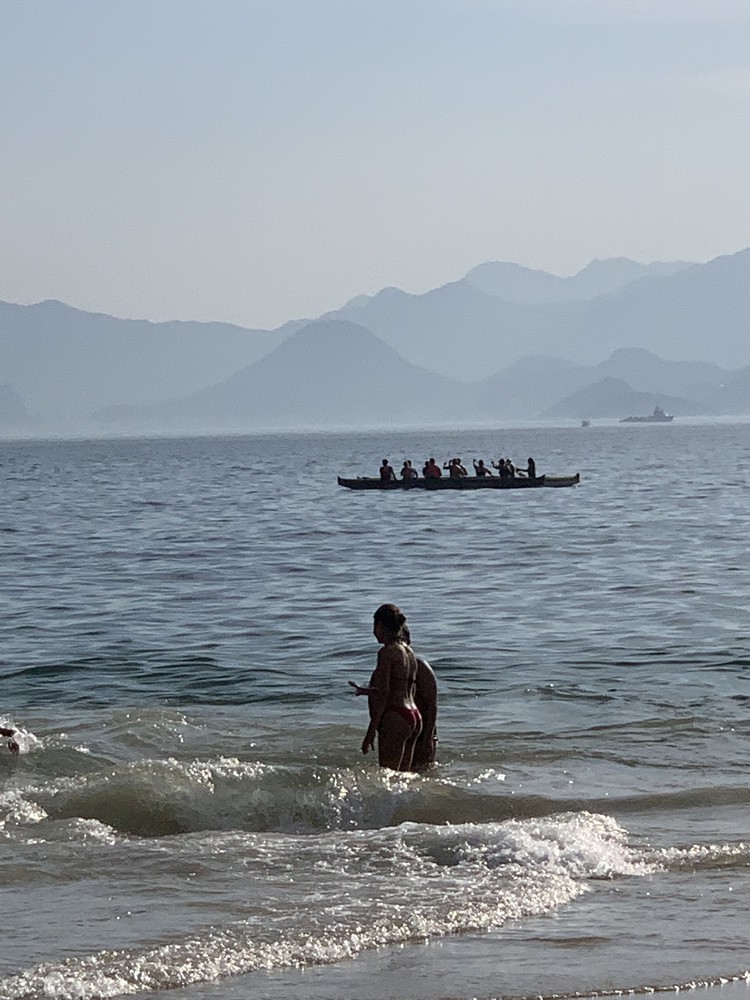}
        \footnotesize{Input}
    \end{minipage}
    \begin{minipage}{0.16\hsize}
        \centering
        \includegraphics[keepaspectratio, width=\hsize]{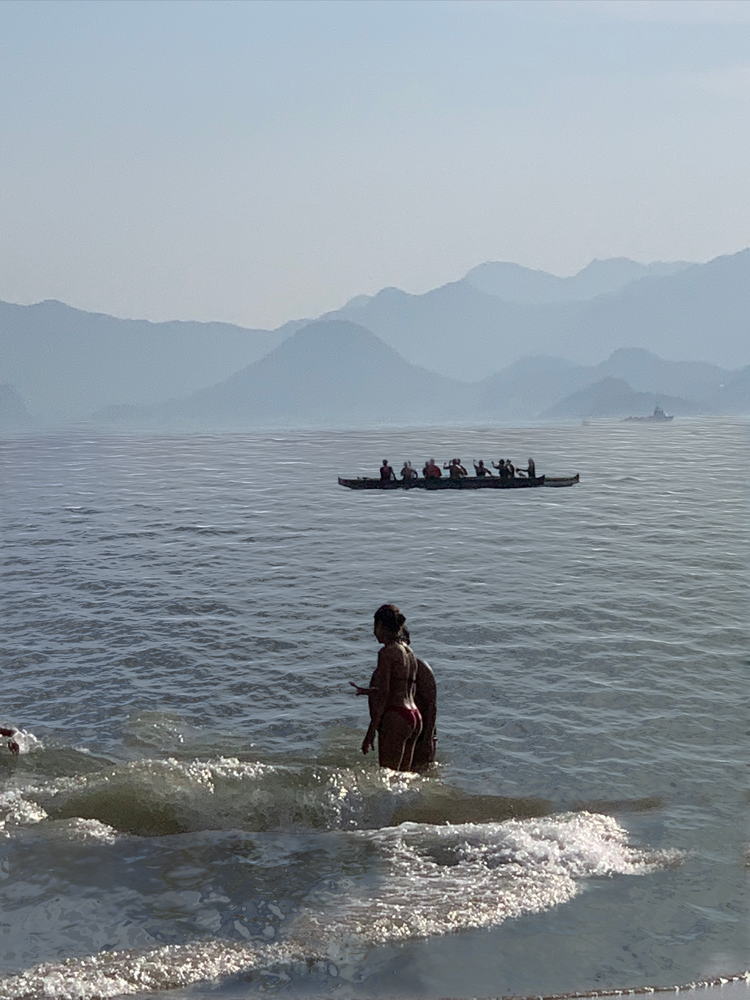}
        \footnotesize{Output}
    \end{minipage}
   \begin{minipage}{0.32\hsize}
        \centering
        \includegraphics[keepaspectratio, width=\hsize]{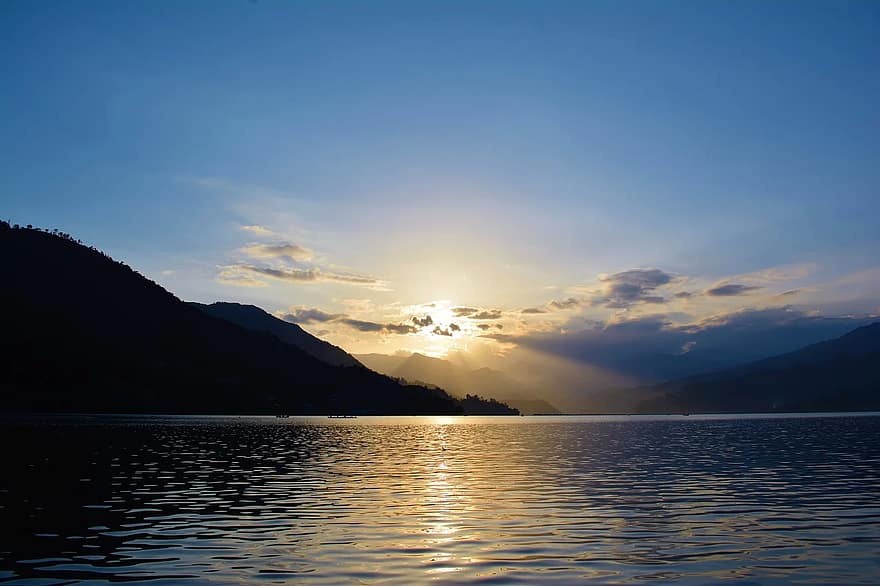}
        \footnotesize{Input}
    \end{minipage}
    \begin{minipage}{0.32\hsize}
        \centering
        \includegraphics[keepaspectratio, width=\hsize]{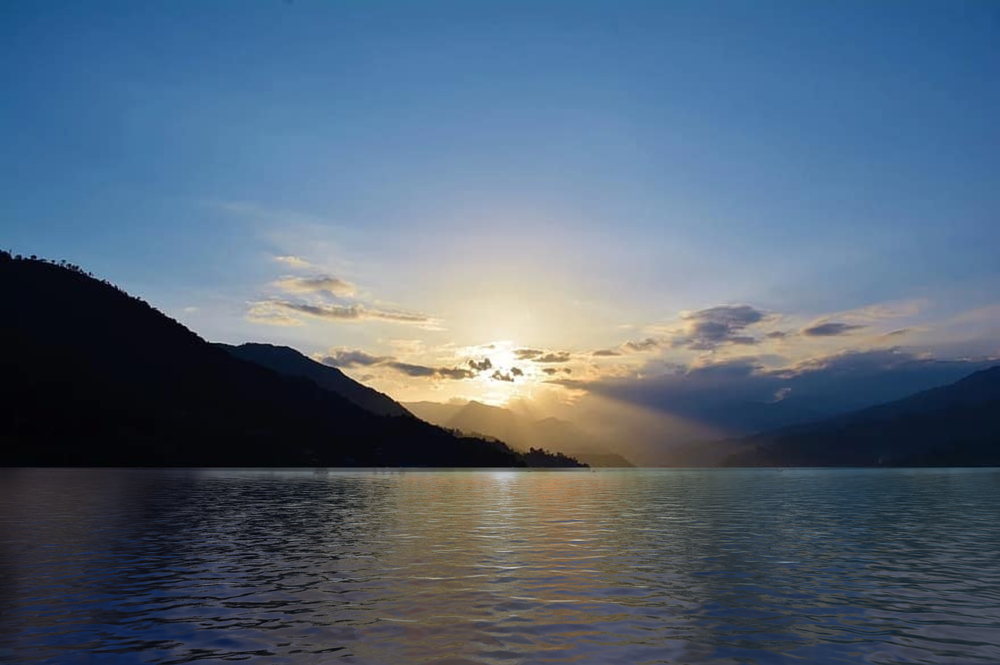}
        \footnotesize{Output}
    \end{minipage}
    
    \begin{tabular}{C{0.29\linewidth}C{0.62\linewidth}}
    (a) & (b)\\
    \end{tabular}
    \caption{Limitations. In (a), our method fails to capture the dynamics of breaking waves. In (b), our method does not handle strong reflection of sunlight. The second input image: Eshan Chandra/Pixabay.}
    \label{fig:failure}
\end{figure}

While our method works on a wide variety of scenes, there are situations where our approach is not applicable due to the inherent limitations of the water surface simulation model employed. For instance, our method cannot simulate flowing water such as waterfalls and dynamic water breaks such as ocean waves on the coastal line (Fig.~\ref{fig:failure}a). We also do not propose a detailed treatment of refraction but instead use a single color in place of the refraction component, which is a valid assumption case of deep water. In addition, our lighting model does not handle strong reflection of sunlight in some cases (Fig.~\ref{fig:failure}b) because both input and output of the reflection texture prediction are standard 24-bit RGB images, which cannot store high radiance by strong lights. Furthermore, we assume that there is one set of parameters for the body of water, and this assumption may not hold very close to the coastline. Our current implementation does not support multiple bodies of water in the same scene, but this can be easily addressed with a trivial extension.

In summary, we present a technique that renders realistic animated water surfaces given a single still input photograph. Our approach determines rendering parameters and water reflection textures using a combination of neural networks and optimization techniques. The results are then fed to our renderer which displays the animated water in real-time. The entire process is fully automatic and relies on a single input image. Our approach generates realistic results for a wide variety of natural scenes with different lighting and water surface conditions, yielding particularly good results for deepwater scenes. The generated 3D scenes show the potential to support a variety of interactive applications.

\begin{acks}
We thank the anonymous reviewers for valuable feedback on our
manuscript and the owners of photographs used in our paper.
The authors from Hong Kong were partly supported by RGC including GRF Grant CityU 11216122.
\end{acks}

\bibliographystyle{ACM-Reference-Format}
\bibliography{main}

\end{document}


\title[Water Simulation and Rendering from a Still Photograph - Supplemental Material]{Water Simulation and Rendering from a Still Photograph\\
Supplemental Material}

\author{Ryusuke Sugimoto}
\orcid{0000-0001-5894-0423}
\affiliation{
 \institution{University of Waterloo}
 \city{Waterloo}
 \state{ON}
 \country{Canada}}
\email{rsugimot@uwaterloo.ca}

\author{Mingming He}
\orcid{0000-0002-9982-7934}
\affiliation{
 \institution{Netflix}
 \city{Los Angeles}
 \state{CA}
 \country{USA}}
\email{hmm.lillian@gmail.com}

\author{Jing Liao}
\orcid{0000-0001-7014-5377}
\affiliation{
 \institution{City University\\of Hong Kong}
 \city{Kowloon}
 \country{Hong Kong}
 }
\email{jingliao@cityu.edu.hk}

\author{Pedro V. Sander}
\orcid{0000-0002-0435-9833}
\affiliation{
 \institution{The Hong Kong University\\of Science and Technology}
 \city{Kowloon}
 \country{Hong Kong}
 }
\email{psander@cse.ust.hk}

\maketitle

\begin{table}[t]
  \caption{Parameter set optimized by the cuckoo search algorithm and used by our renderer.}
  \label{tab:parameters}
  \begin{tabular}{ccc}
    \toprule
    \textbf{Parameter}&\textbf{DoF}&\textbf{Range}\\
    \midrule
    Wind speed&1&[1.5, 30.0]\\
    Wind direction&1&[0.0, 180.0]\\
    Wave choppiness&1&[0.0, 3.0]\\
    Camera height&1&[1.0, 75.0]\\
    Camera angle&1&[45.0, 105.0]\\
    Camera field of view&1&[45.0, 90.0]\\
    Water color&3&[0.0, 1.0]\\
    Level 0 SH lighting coefficients&3&[0.0, 2.0]\\
    Level 1 SH lighting coefficients&9&[-1.0, 1.0]\\
    \bottomrule
    \end{tabular}
\end{table}

\section{Dataset for water segmentation}
To train the water segmentation network, we first collect $1,995$ real water images ($1,895$ for training and $100$ for testing) to form a real water image dataset with ground truth annotation for water from multiple public scene datasets, SUN~\cite{DBLP:conf/cvpr/XiaoHEOT10}, ADE20K~\cite{DBLP:conf/cvpr/ZhouZPFB017} and Places~\cite{DBLP:conf/nips/ZhouLXTO14}. However, such a small amount of training data results in overfitting and limited generalization capacity. Thus, we automatically augment the training data using a synthetic water image dataset with $10,000$ randomly selected images from the texture image generation dataset described in Section~\ref{sec:datatex}.
To increase the diversity, we combine either background from the real water image dataset or foreground objects from the COCO dataset~\cite{DBLP:conf/eccv/LinMBHPRDZ14}. When using both datasets, we train the network for the same number of iterations on each of them. By combining the synthetic data with the real data, the segmentation accuracy is largely improved, which can be seen from the mean intersection over union (IOU) metric on the testing dataset, $0.8405$ for the model trained with the synthetic data while $0.8160$ for the model without such augmentation.

\section{Dataset for texture image generation}
\label{sec:datatex}
We use the renderer described in Section 5 of the paper to create the dataset. The parameters are chosen randomly from a valid input range. The water surface color is chosen uniformly randomly from a valid subspace of HSV color space and then converted to RGB space. This limits the color range to be ``blueish'' while avoiding overly bright or overly vivid colors. The spherical harmonics constants are used to approximate the surrounding global lighting effects. We randomly generate a set of colors for several points on a unit hemisphere and use them to generate SH constants. Thus, we can generate infinitely many combinations of SH constants to ensure enough variety in lightning. In place of the reflection texture, we use an image that is flipped upside-down. Even though the reflection on water surface should be distorted in real scenes, the flipped image is a good approximation. We expect that what the reflection texture generation network should learn is irrelevant to such distortions. Based on this assumption, we chose images to be used as reflection textures from a subset of the Places database~\cite{DBLP:conf/nips/ZhouLXTO14} with outdoor scenes and sky images from the SWINySEG\cite{Dev2019} dataset. In our dataset generation, we further made a simplification of the model. We assumed that the water region boundary is a straight line that lies within a random distance from the patch in image coordinates. This is required because each reflection image patch is supposed to be entirely in a water area and we do not need any information about the objects outside of the image patch.

\section{Details of texture image generation network}
As mentioned in the paper, the network architecture is based on UNet with residual blocks. The input patches are of size $224 \times 224$. The encoder consists of series of convolution layers with padding: $3 \times 3$ with stride 1, $7 \times 7$ with stride 2, $3 \times 3$ with stride 2, and $3 \times 3$ with stride 2; except the last convolution, each  followed by batch normalization and a ReLU. There are five intermediate residual blocks which are followed by batch normalization and a ReLU. The decoder contains three upsampling layers of factor two. The first two upsampling layers are respectively followed by concatenation of feature map with the corresponding feature from the encoder, convolution, batch normalization, and a ReLU. The last upsampling layer is followed by feature concatenation and convolution layers. The convolution kernels in the decoder are $1 \times 1$, $3 \times 3$, and $1 \times 1$, in sequential order, all with stride 1.

\section{Parallelized Cuckoo Search}
\begin{figure}[h]
\includegraphics[keepaspectratio, width=0.49\hsize]{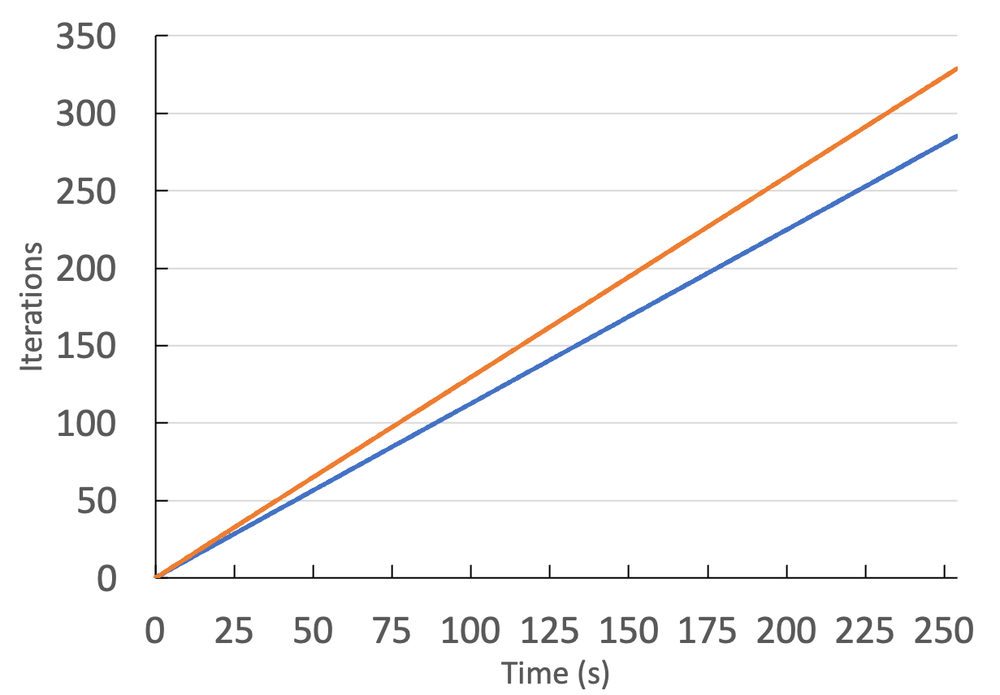}
\includegraphics[keepaspectratio, width=0.49\hsize]{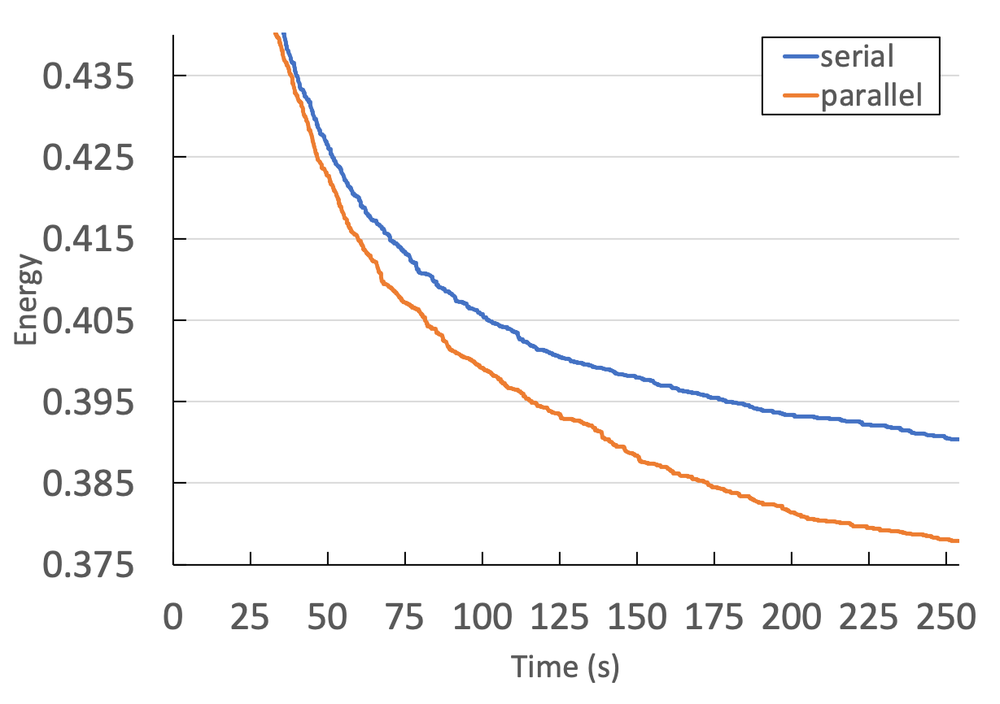}
\caption{Comparison of the serial and parallel cuckoo search algorithms. All figures show average curves for 55 test images. On average, the termination condition is met after 269 iterations (212 seconds) with an average energy of 0.389.}\label{fig:cuckoo_parallel}
\end{figure}

The energy computation is the most expensive component of our pipeline. It can be partitioned into frame rendering using the candidate parameter set and the DISTS and HSV color histogram energy evaluation. Rendering and DISTS evaluation accounts the majority of execution time. To improve efficiency, we perform evaluation for $k-1$th iteration and rendering for $k$th iteration concurrently (Algorithm \ref{alg:parallel_cuckoo}). The concurrent execution of rendering and energy evaluation allows us to maximize the GPU utilization from 80-90\% to 98\%, and allows more energy evaluation per unit time. While the optimization process is not completely equivalent to the original serial version because of its lazy update feature, our experiments validate the effectiveness of the method (Figure \ref{fig:cuckoo_parallel}). At the begging of evaluation, there is no significant difference of the energy curve due to the lazy update of candidate solutions. However, at later iterations, the difference in number of energy evaluations per unit time between the serial implementation and the parallel implementation lead to the difference of energy curves.

\begin{algorithm}[t]
    \caption{Cuckoo Search}
    \label{alg:cuckoo}

    \While {termination condition not met}{
    \ForAll {nest $n_i$}{
        Get a cuckoo egg $\mathbf{x}_i'$ by L\'{e}vy flight from $\mathbf{x}_i$.\\
        Choose a random nest $n_j$ to lay the egg.\\
        \If {$Score(\mathbf{x}_i') < Score(\mathbf{x}_j')$}
            {replace $\mathbf{x}_j$ with $\mathbf{x}_i'$. }
    }
    \ForAll {nest $n_i$} {
        Get a cuckoo egg $\mathbf{x}_i'$ by mutation from $\mathbf{x}_i$\\
        \If {$Score(\mathbf{x}_i') < Score(\mathbf{x}_i)$}
           {replace $\mathbf{x}_i$ with $\mathbf{x}_i'$.}
    }
    Replace the worst $k$ eggs with randomly generated eggs.
    }
    \Return $\mathbf{x}_{best}$
\end{algorithm}

\begin{algorithm}[t]
    \caption{Parallel Cuckoo Search ($k$th iteration)}
    \label{alg:parallel_cuckoo}

    solutions\_l\'{e}vy$_k\leftarrow$ generate\_solutions\_l\'{e}vy(curr\_nests)\\
    solutions\_mut$_k\leftarrow$ generate\_solutions\_mut(curr\_nests)\\
    solutions\_rand$_k\leftarrow$ generate\_solutions\_rand(curr\_nests)\\
    ~\\
    frames$_{k-1} \leftarrow$ receive\_rendered\_frames()\\
    submit\_rendering\_tasks($[$solutions\_l\'{e}vy$_k$, solutions\_mut$_k$,\\
    \quad\quad solutions\_rand$_k]$ )\\
    $[$scores\_l\'{e}vy$_{k-1}$,  scores\_mut$_{k-1}$, scores\_rand$_{k-1}]$\\
    \quad\quad $\leftarrow$ evaluate\_energy\_function(frames$_{k-1}$)\\
    update\_nests\_l\'{e}vy(solutions\_l\'{e}vy$_{k-1}$, scores\_l\'{e}vy$_{k-1}$)\\
    update\_nests\_mut(solutions\_mut$_{k-1}$, scores\_mut$_{k-1}$)\\
    update\_nests\_rand(solutions\_rand$_{k-1}$, scores\_rand$_{k-1}$)\\

\end{algorithm}

\begin{figure}[t]
    \centering
    \begin{minipage}{0.32\hsize}
        \centering
        \includegraphics[keepaspectratio, width=\hsize]{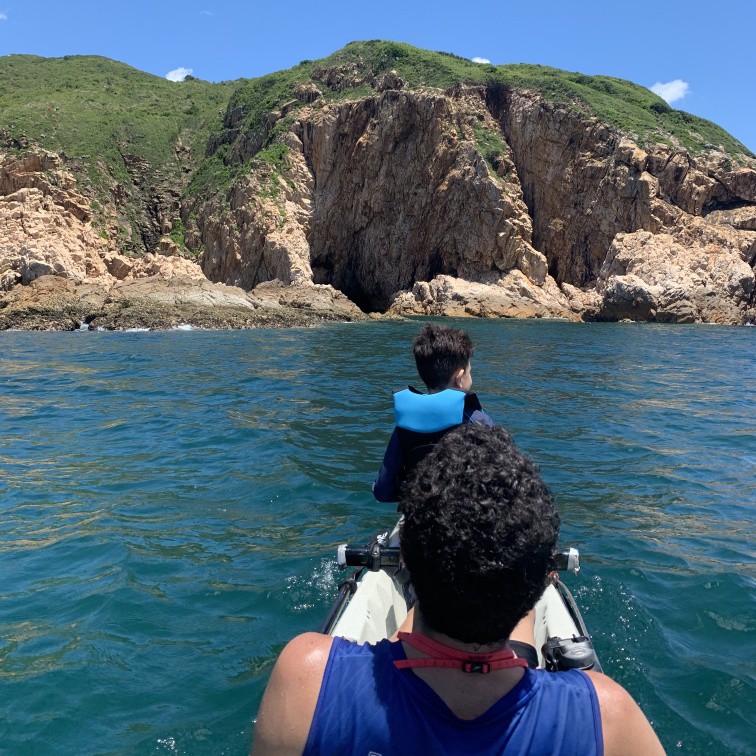}
    \end{minipage}
    \begin{minipage}{0.32\hsize}
        \centering
        \includegraphics[keepaspectratio, width=\hsize]{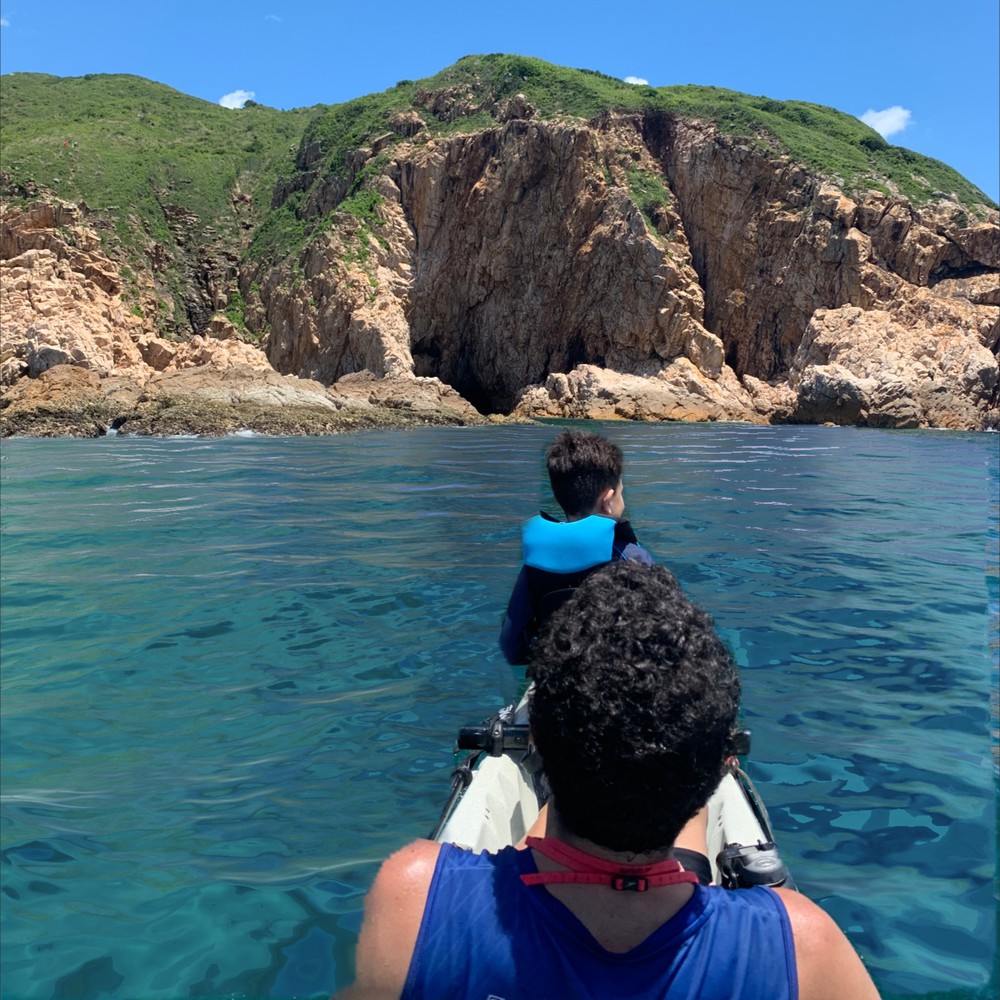}
    \end{minipage}
    \begin{minipage}{0.32\hsize}
        \centering
        \includegraphics[keepaspectratio, width=\hsize]{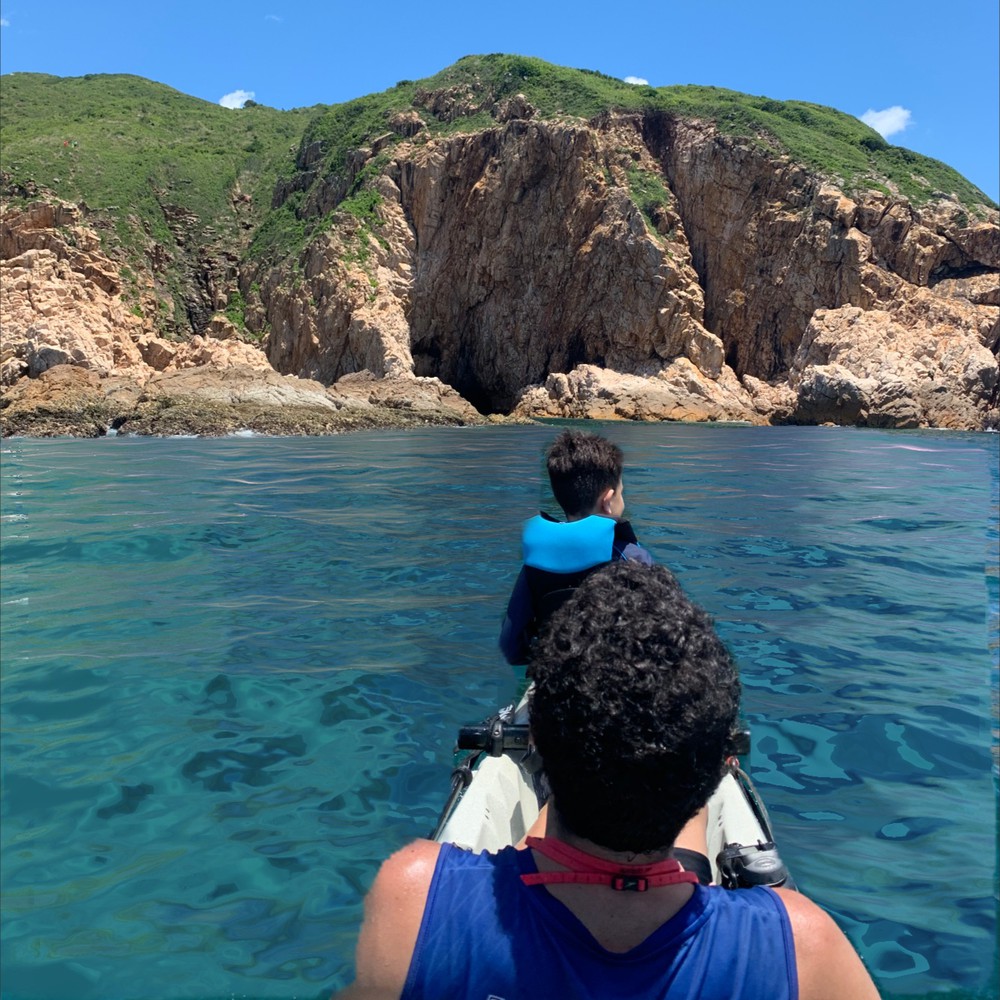}
    \end{minipage}
    \centering
    \begin{minipage}{0.32\hsize}
        \centering
        \includegraphics[keepaspectratio, width=\hsize]{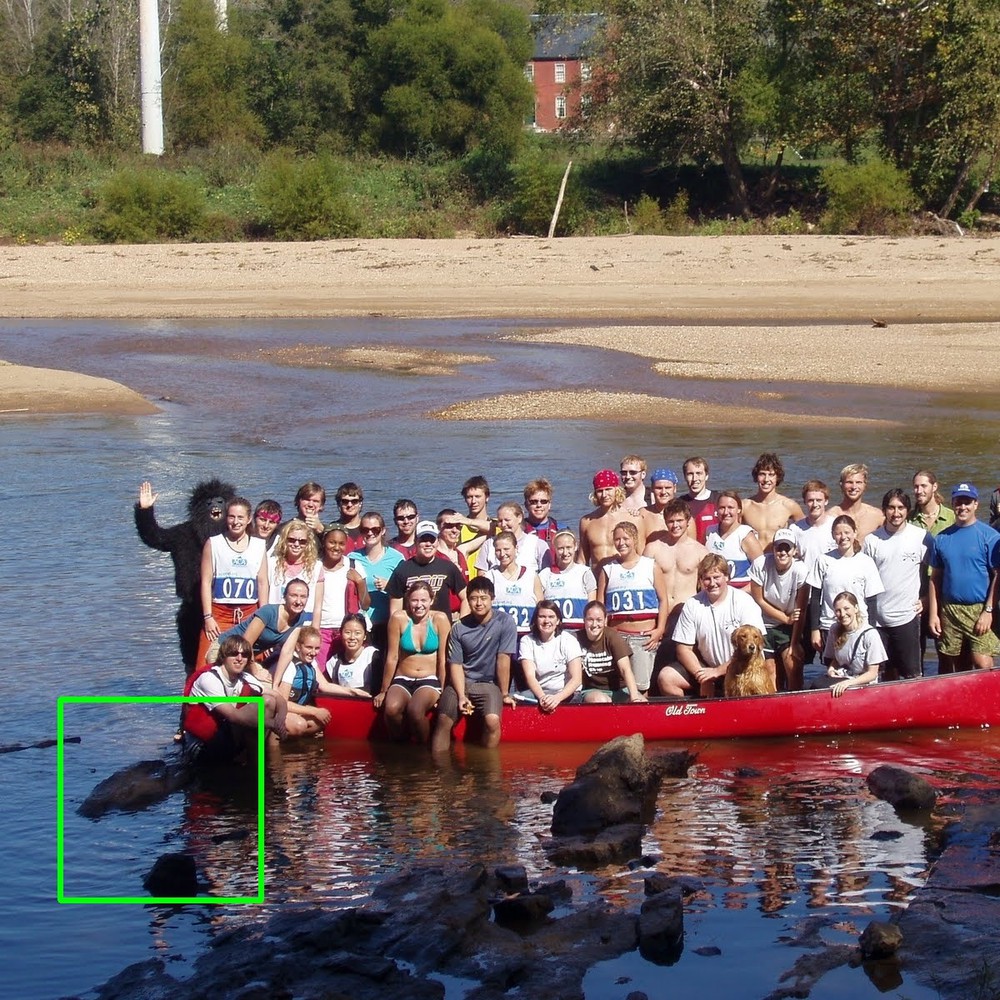}
        \subcaption{input}
    \end{minipage}
    \begin{minipage}{0.32\hsize}
        \centering
        \includegraphics[keepaspectratio, width=\hsize]{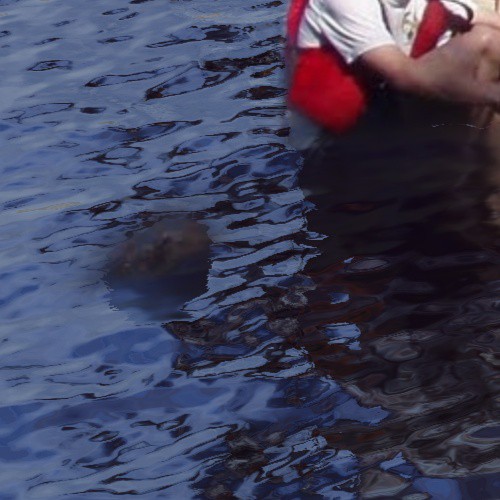}
        \subcaption{w/o optimization}
    \end{minipage}
    \begin{minipage}{0.32\hsize}
        \centering
        \includegraphics[keepaspectratio, width=\hsize]{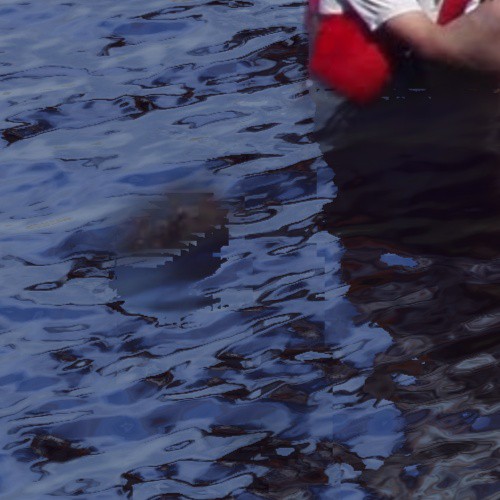}
        \subcaption{with optimization}
    \end{minipage}
    \caption{In most cases, such as in the top row, there is no sacrifice in quality when performing the reflection computation optimization. However, in some complex boundaries, such as in the bottom row, some artifacts can be noticed. The second input image is from Places~\cite{DBLP:conf/nips/ZhouLXTO14}.}
    \label{fig:renderer_optim_success}
\end{figure}

\section{Influence of Initial Parameters in Cuckoo Search}
\begin{figure}[t]
    \centering
    \begin{minipage}{0.49\hsize}
        \centering
        \includegraphics[keepaspectratio, width=\hsize]{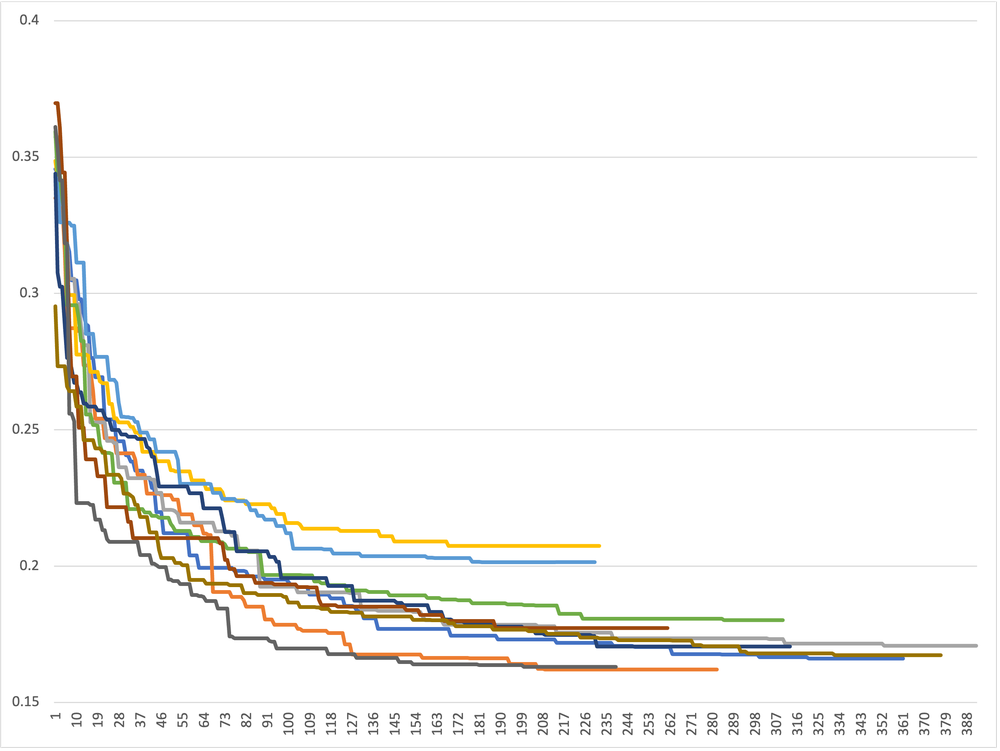}
    \end{minipage}
    \begin{minipage}{0.49\hsize}
        \centering
        \includegraphics[keepaspectratio, width=\hsize]{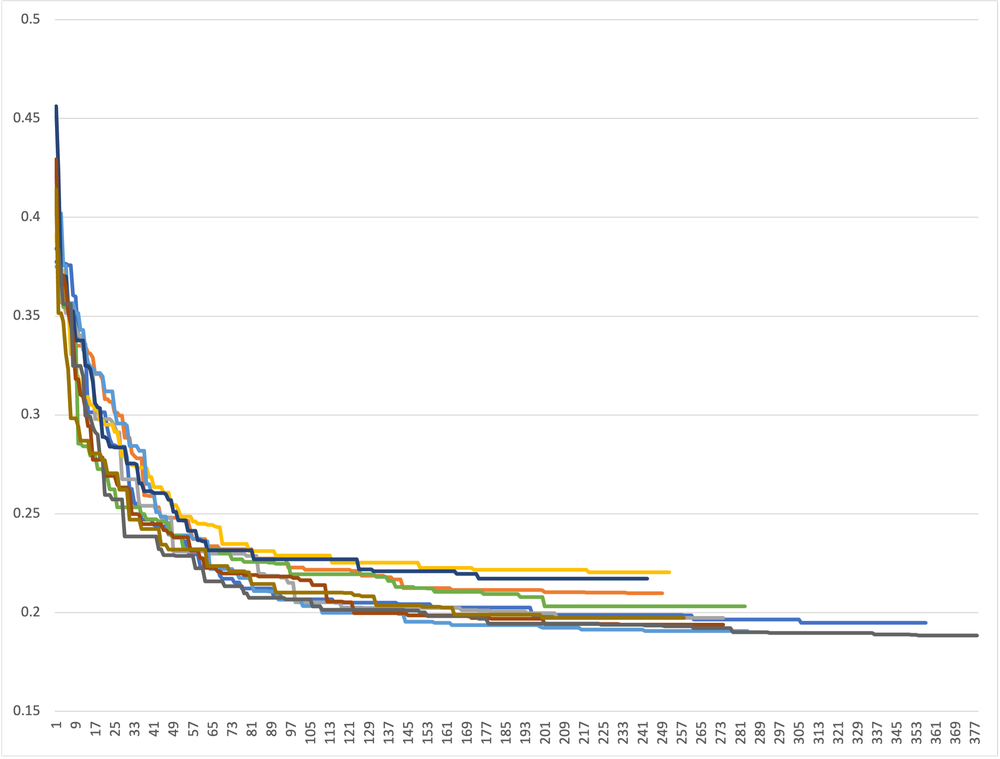}
    \end{minipage}
    \caption{Optimization processes for the scenes in Fig. 1 and Fig. 8 first row with 10 different initial parameter sets, respectively. The vertical axis represents the energy, and the horizontal axis represents the number of iterations. Though the optimization process depends on the initial parameter set, they converge after enough iterations.}
    \label{fig:optim_process}
\end{figure}

\begin{figure}[t]
    \centering
    \begin{minipage}{0.32\hsize}
        \centering
        \includegraphics[keepaspectratio, width=\hsize]{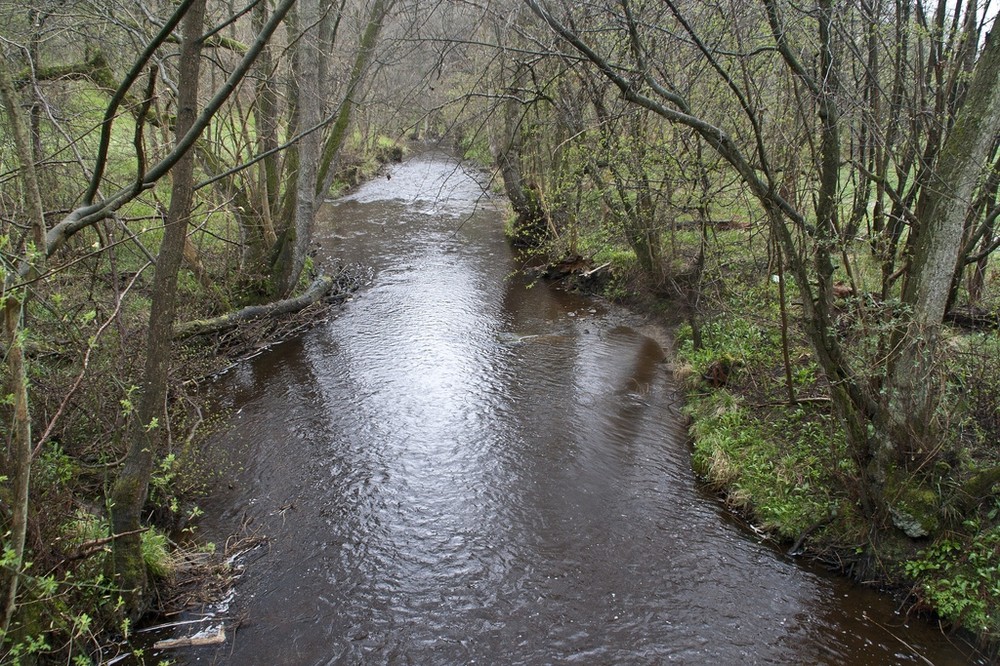}
    \end{minipage}
    \begin{minipage}{0.32\hsize}
        \centering
        \includegraphics[keepaspectratio, width=\hsize]{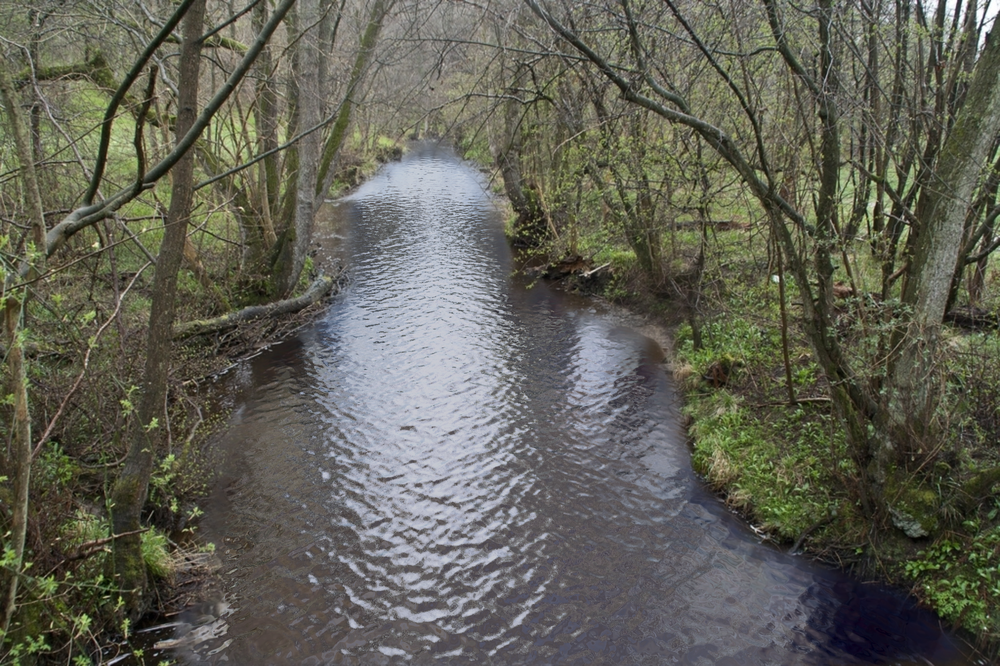}
    \end{minipage}
    \begin{minipage}{0.32\hsize}
        \centering
        \includegraphics[keepaspectratio, width=\hsize]{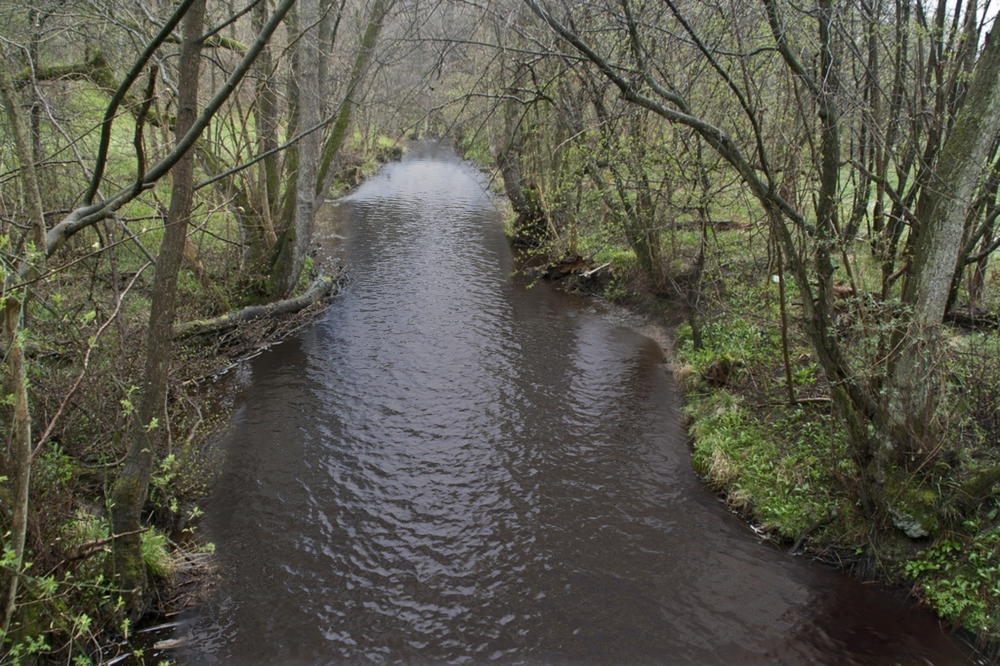}
    \end{minipage}
    \centering
    \begin{minipage}{0.32\hsize}
        \centering
        \includegraphics[keepaspectratio, width=\hsize]{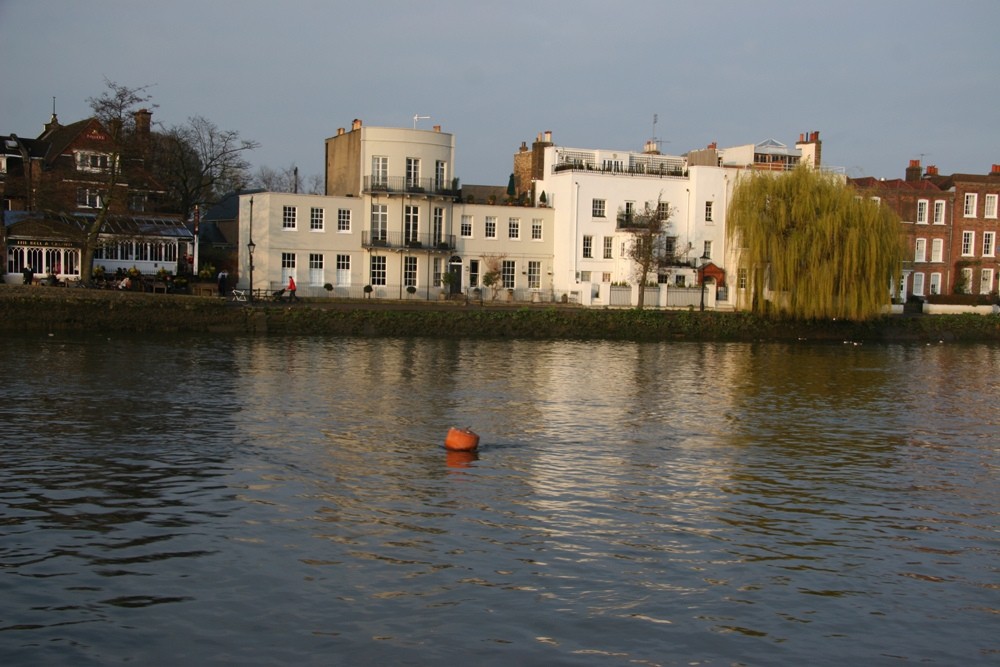}
        \subcaption{input}
    \end{minipage}
    \begin{minipage}{0.32\hsize}
        \centering
        \includegraphics[keepaspectratio, width=\hsize]{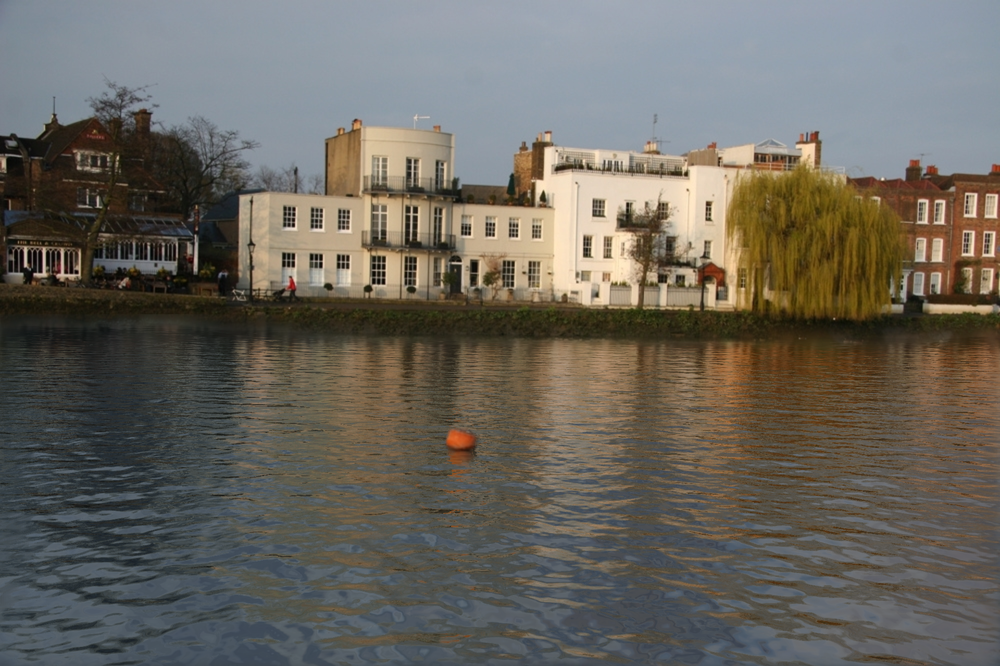}
        \subcaption{best result}
    \end{minipage}
    \begin{minipage}{0.32\hsize}
        \centering
        \includegraphics[keepaspectratio, width=\hsize]{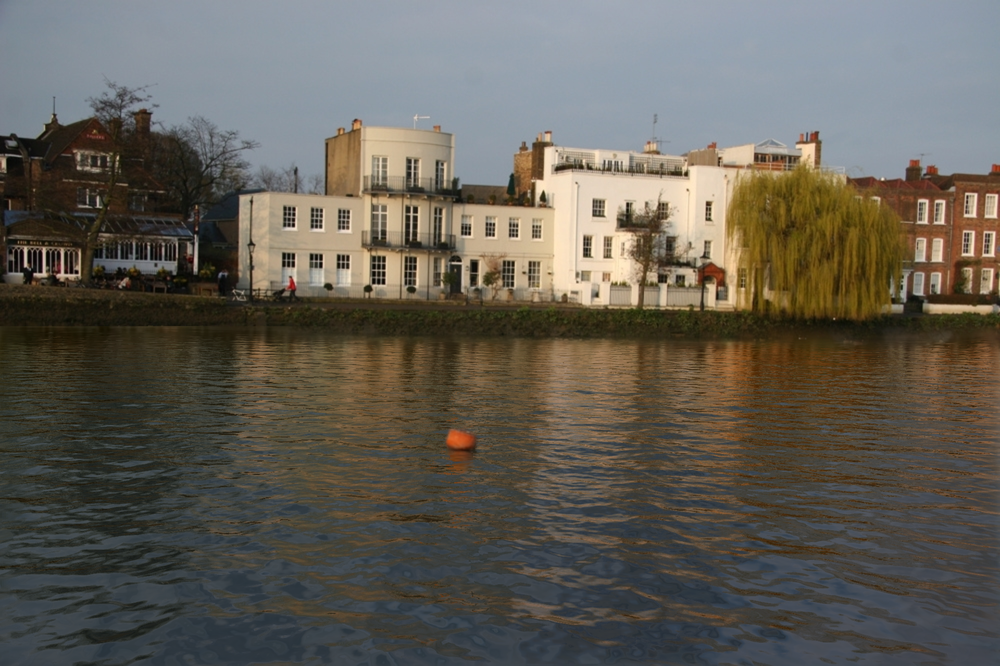}
        \subcaption{worst result}
    \end{minipage}
    \caption{Comparison of the best and worst optimization results from Fig. ~\ref{fig:optim_process}. Our termination criterion is designed conservative enough and even the optimization processes that terminated with highest (worst) scores gives sets of parameters that give us plausible renderings. Input images: Edward Nicholl/Flickr (top) and - Paul -/Flickr (bottom).}
    \label{fig:optim_results}
\end{figure}

Because of the random nature of the search, the optimization process does depend on the initial candidate parameter set. However, we set a conservative enough threshold as the termination condition to let the optimization process reach small enough energy so that it will not give us a noticeable visual difference with respect to the initial candidate parameter set (see Fig.~\ref{fig:optim_process} and Fig.~\ref{fig:optim_results}). We can choose an even stricter termination condition if desired.

\section{Study on the Design of the Energy Function}
\begin{figure*}[t]
    \centering
    \begin{minipage}{0.205\hsize}
        \centering
        \includegraphics[keepaspectratio, width=\hsize]{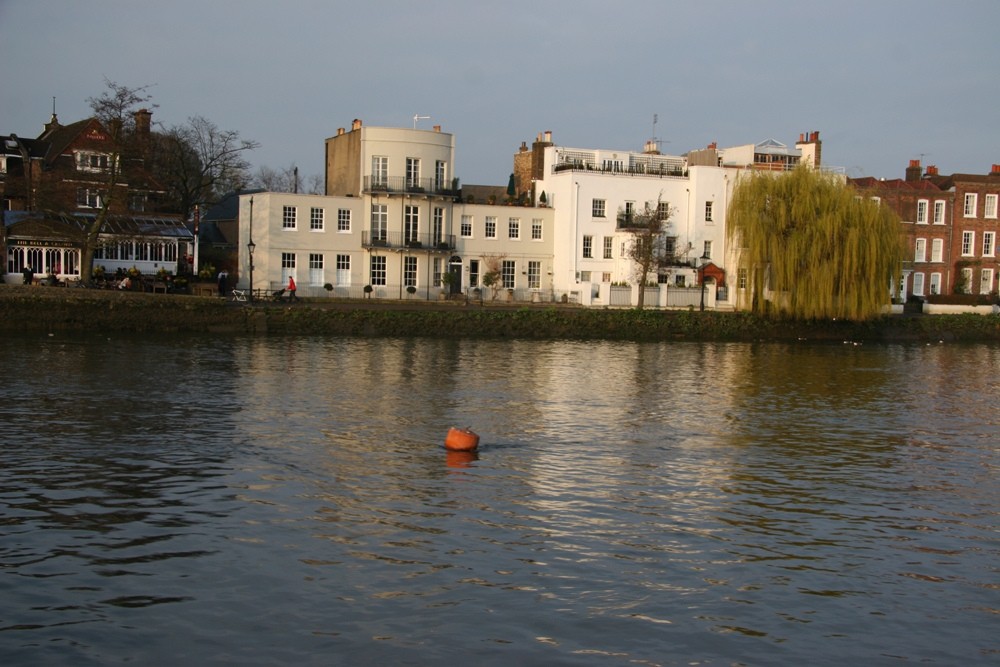}
        \includegraphics[keepaspectratio, width=\hsize]{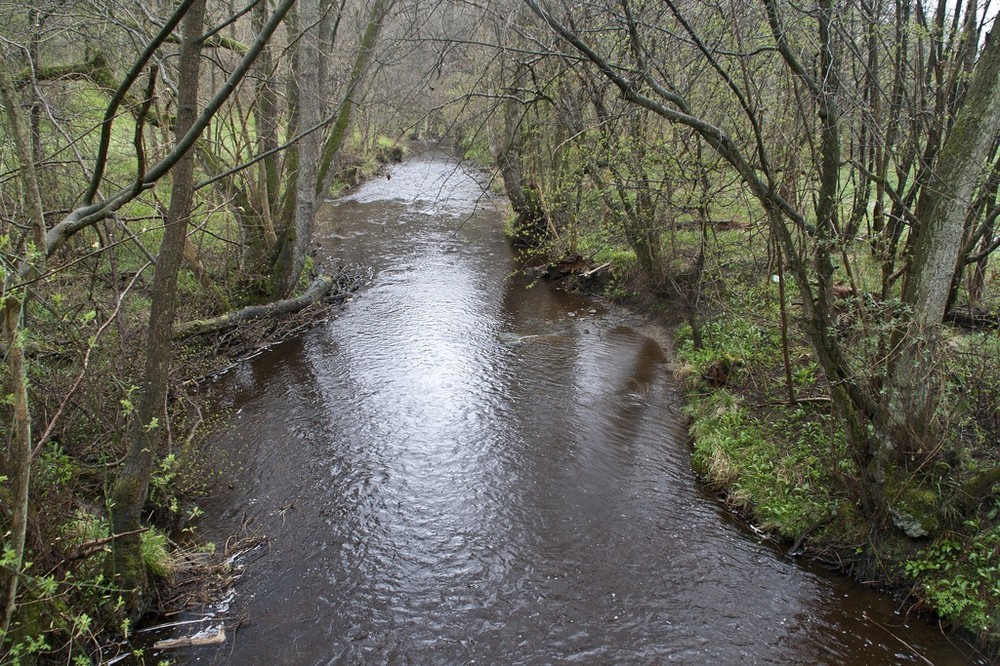}
        \includegraphics[keepaspectratio, width=\hsize]{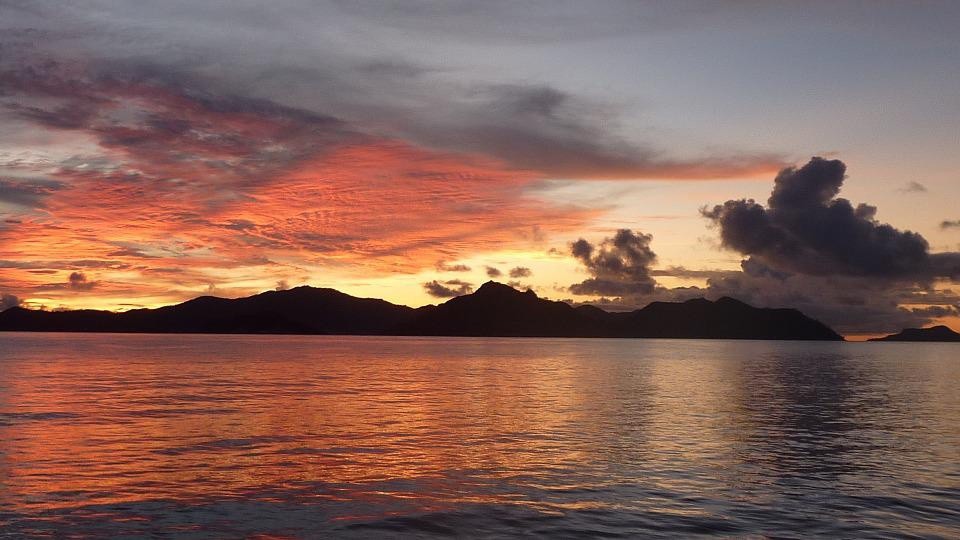}
        \includegraphics[keepaspectratio, width=\hsize]{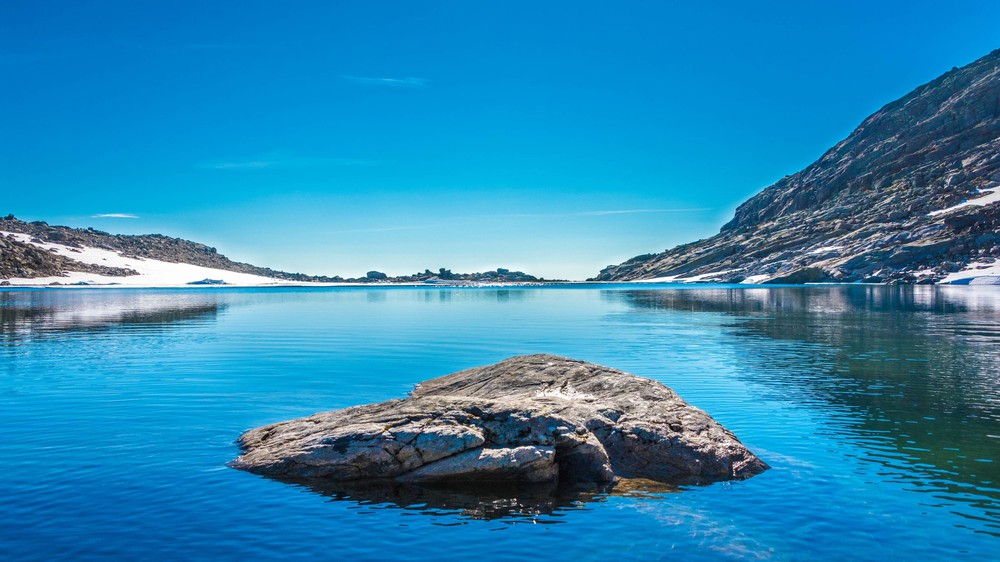}
        \includegraphics[keepaspectratio, width=\hsize]{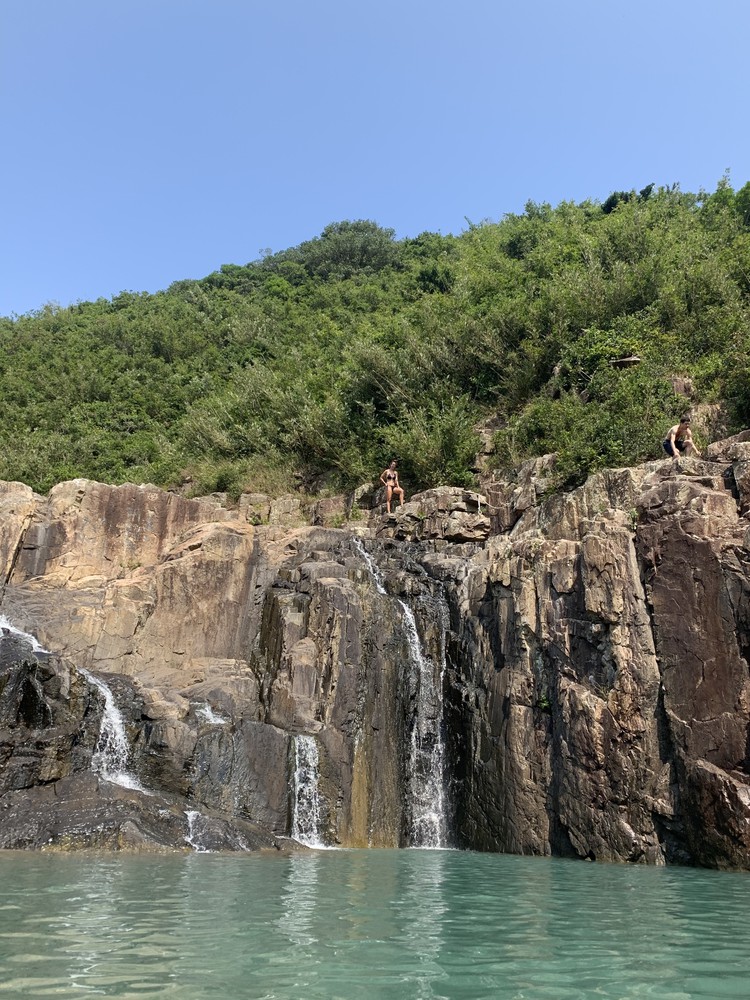}
        \includegraphics[keepaspectratio, width=\hsize]{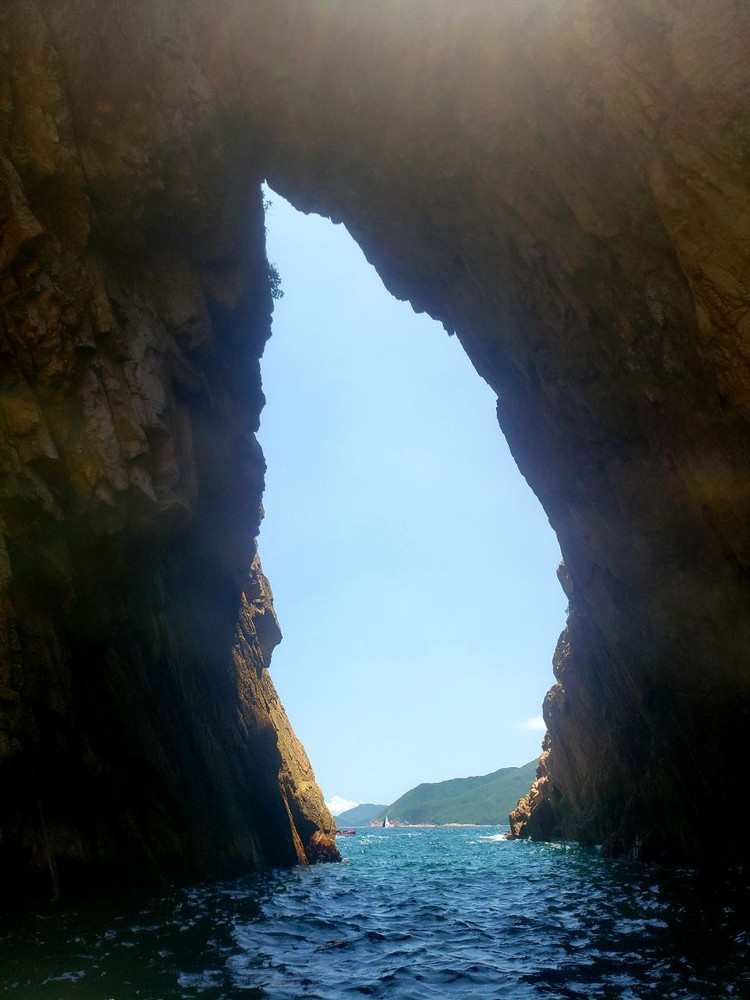}
    \end{minipage}
    \begin{minipage}{0.205\hsize}
        \centering
        \includegraphics[keepaspectratio, width=\hsize]{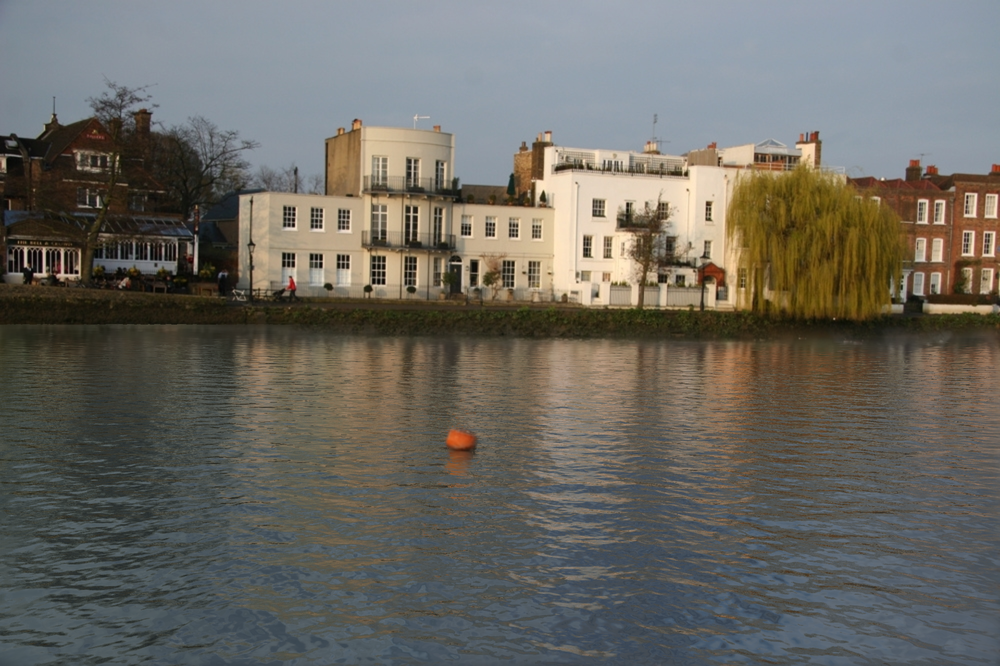}
        \includegraphics[keepaspectratio, width=\hsize]{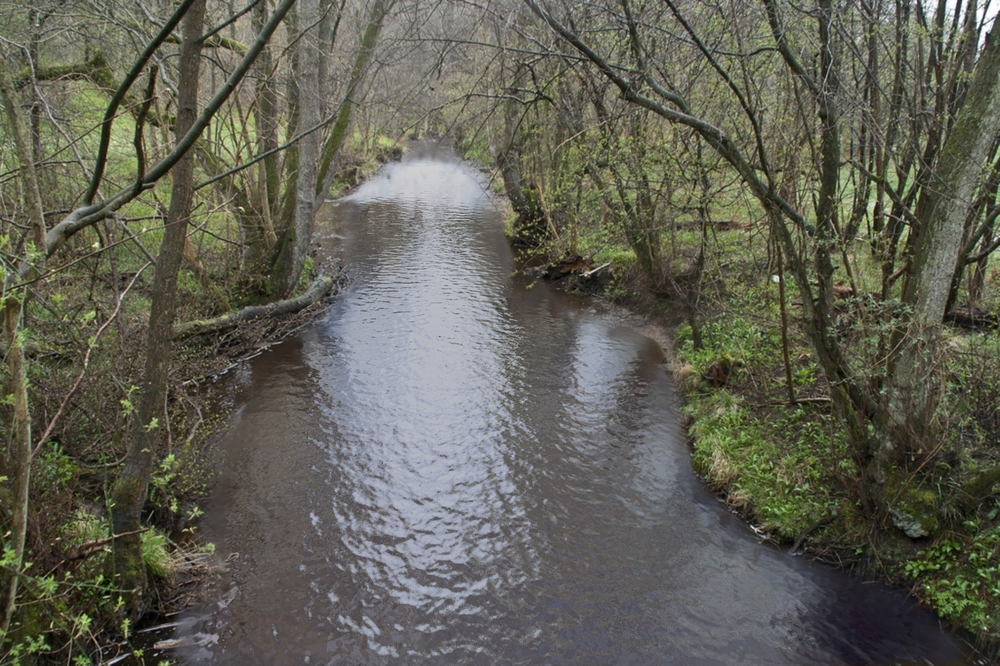}
        \includegraphics[keepaspectratio, width=\hsize]{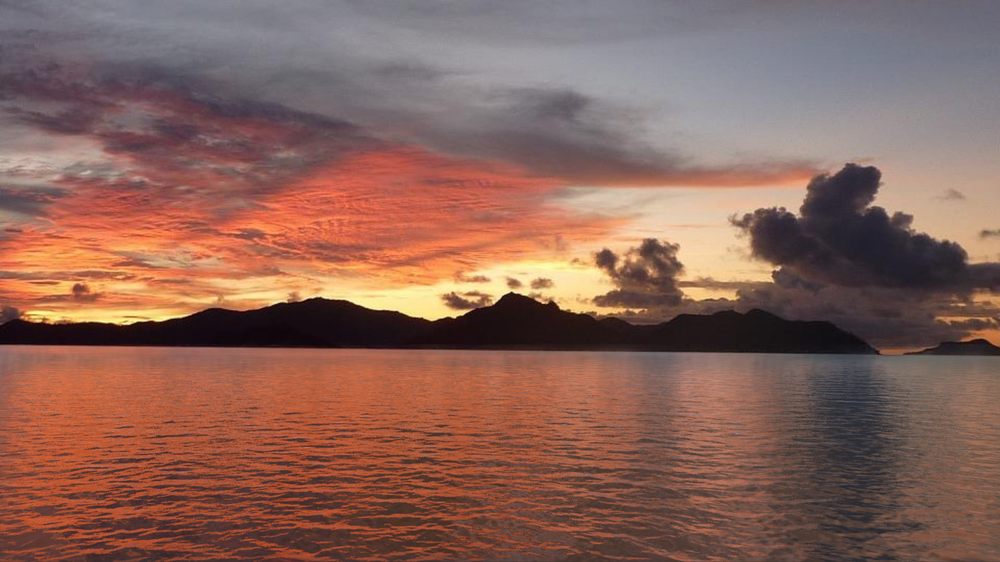}
        \includegraphics[keepaspectratio, width=\hsize]{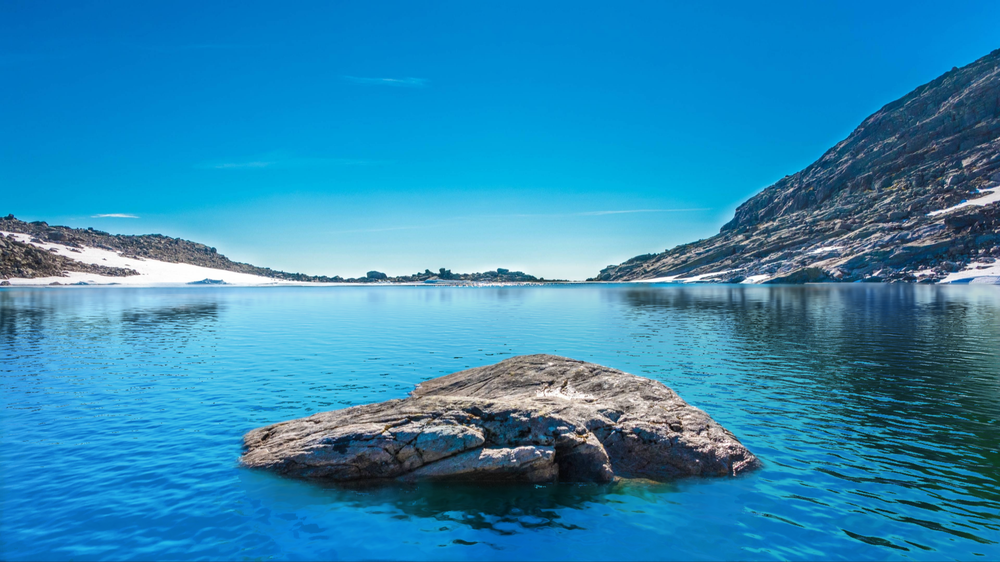}
        \includegraphics[keepaspectratio, width=\hsize]{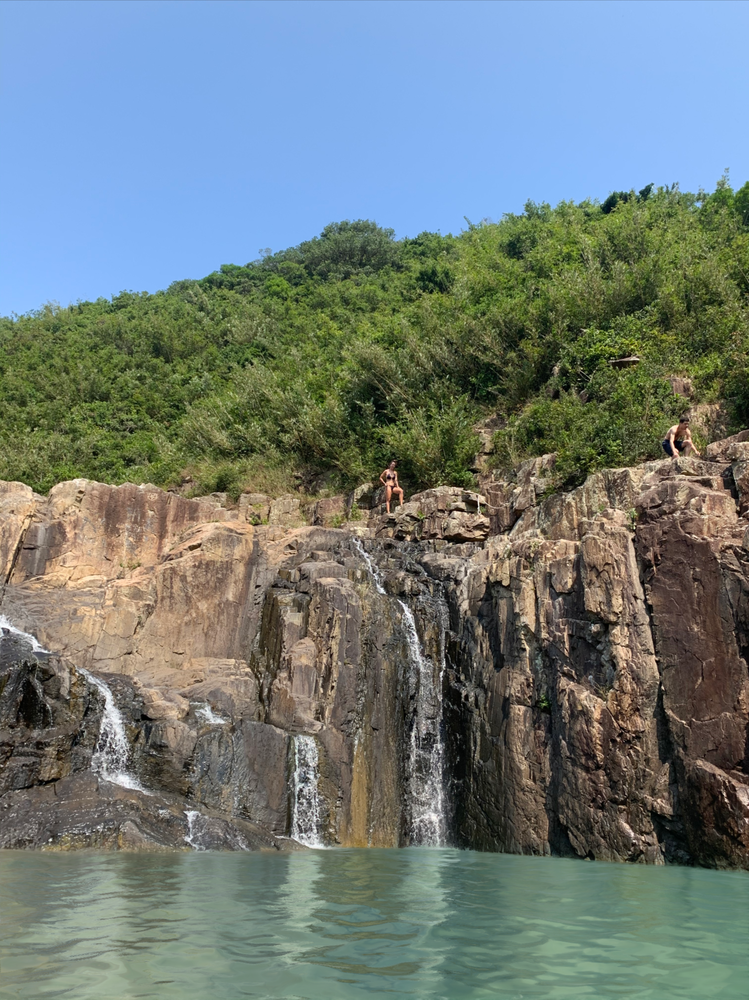}
        \includegraphics[keepaspectratio, width=\hsize]{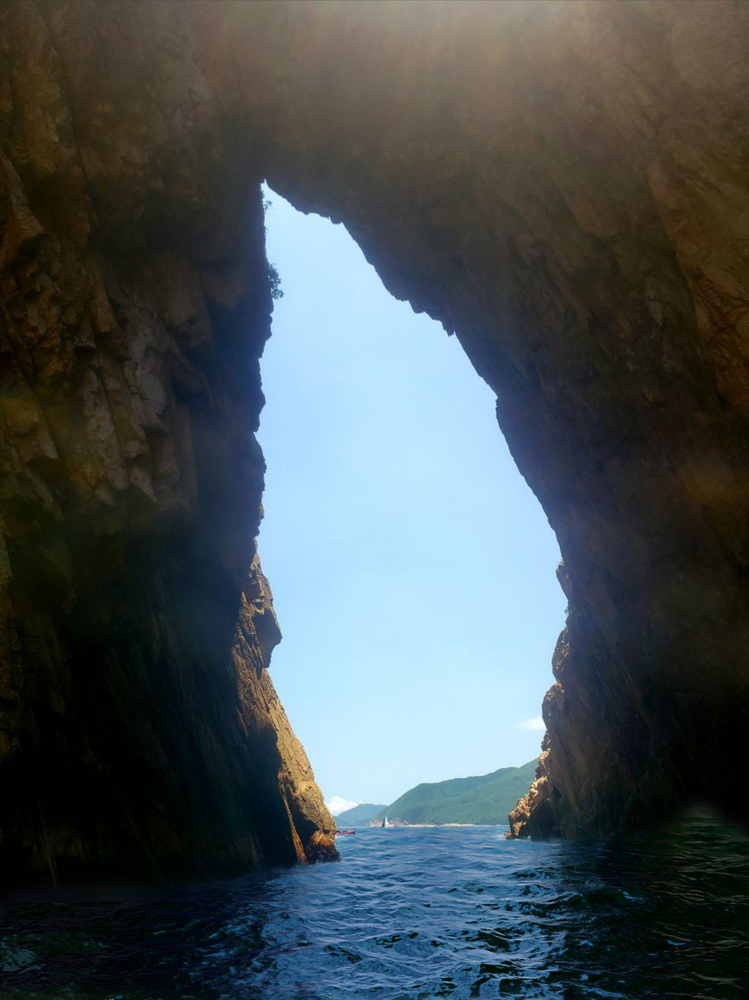}
    \end{minipage}
    \begin{minipage}{0.205\hsize}
        \centering
        \includegraphics[keepaspectratio, width=\hsize]{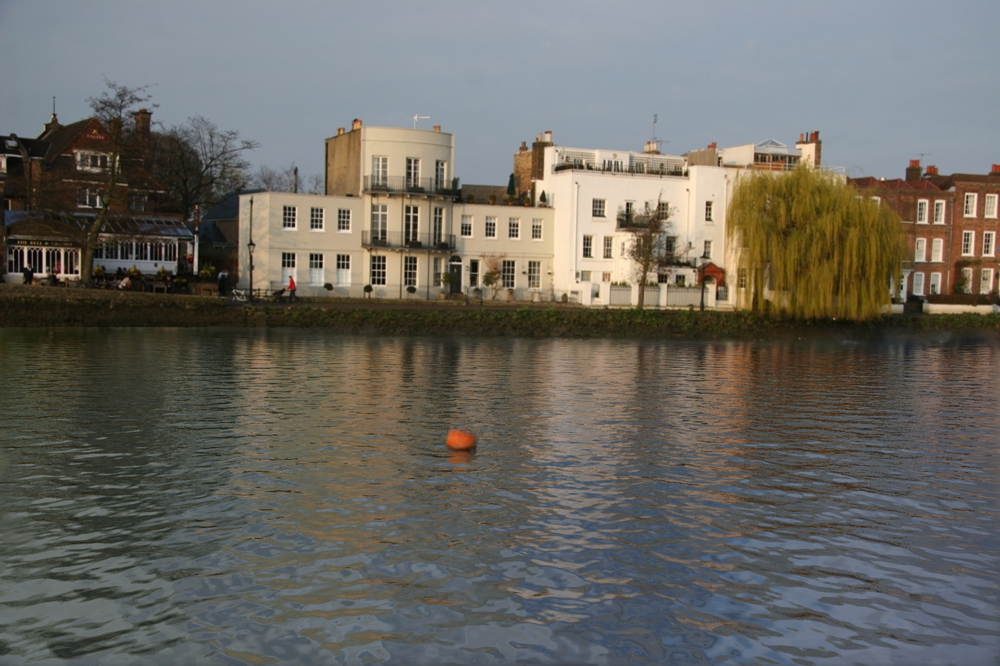}
        \includegraphics[keepaspectratio, width=\hsize]{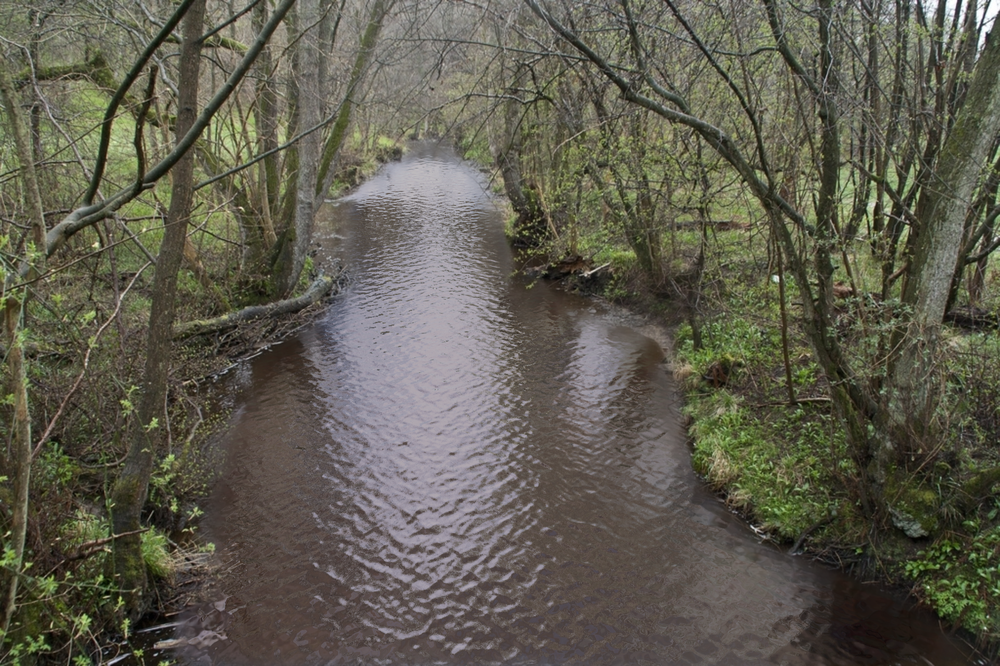}
        \includegraphics[keepaspectratio, width=\hsize]{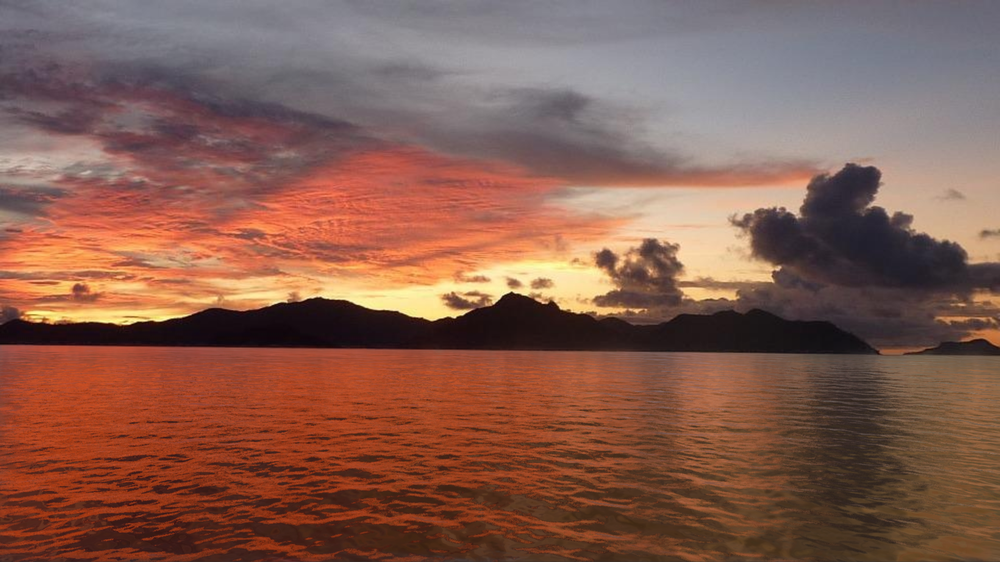}
        \includegraphics[keepaspectratio, width=\hsize]{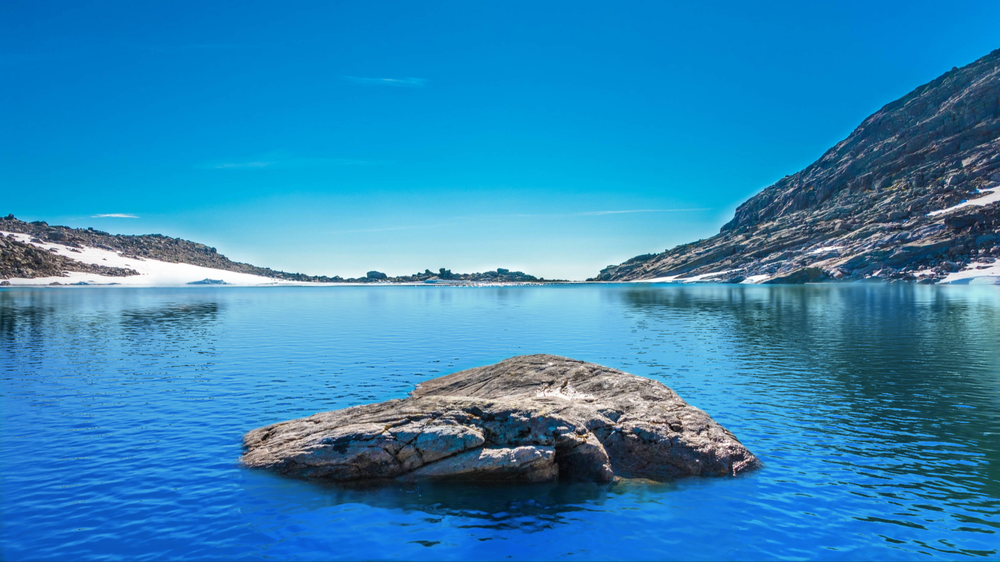}
        \includegraphics[keepaspectratio, width=\hsize]{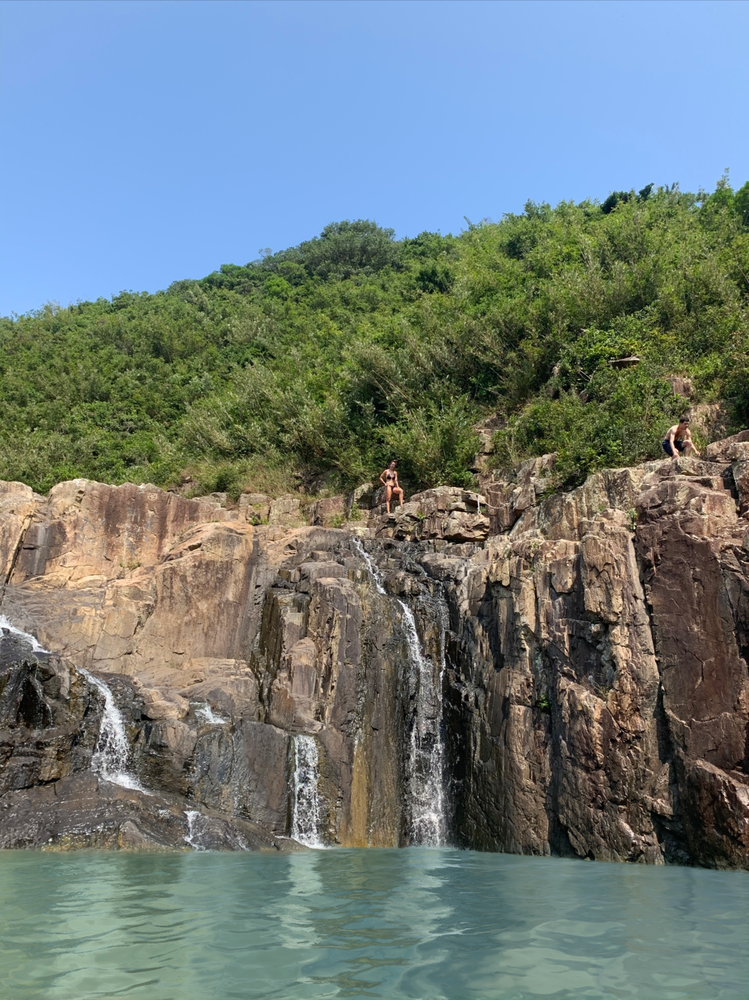}
        \includegraphics[keepaspectratio, width=\hsize]{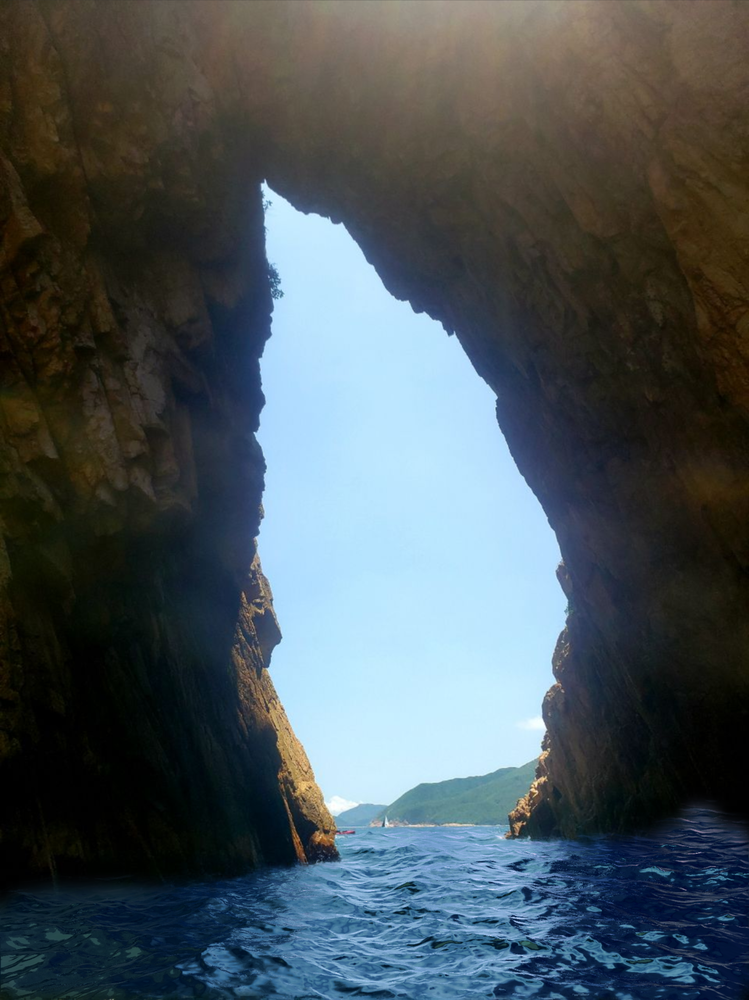}
    \end{minipage}
      \begin{minipage}{0.205\hsize}
        \centering
        \includegraphics[keepaspectratio, width=\hsize]{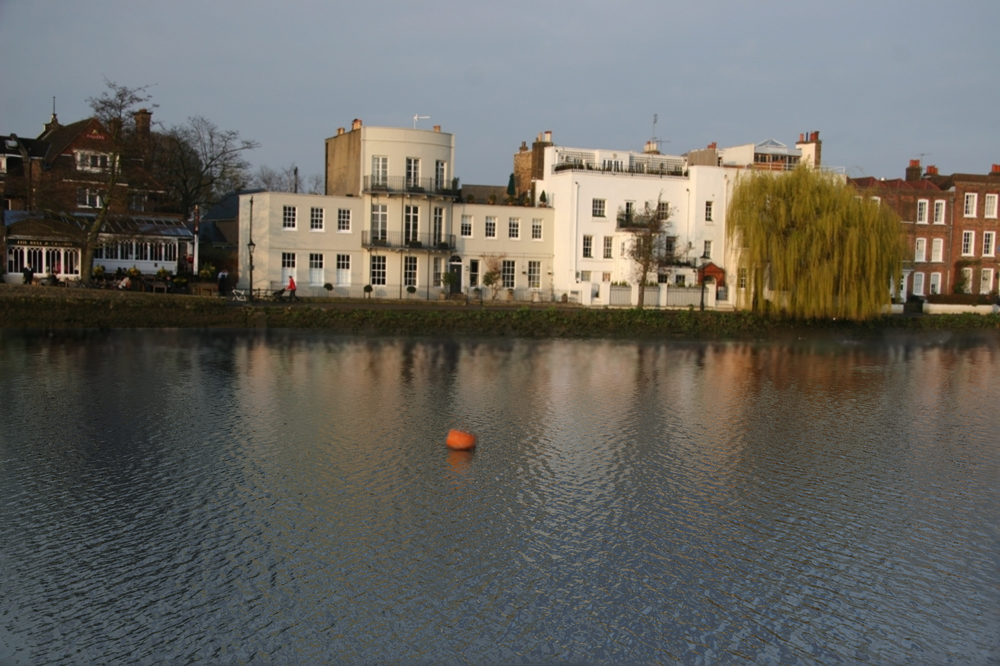}
        \includegraphics[keepaspectratio, width=\hsize]{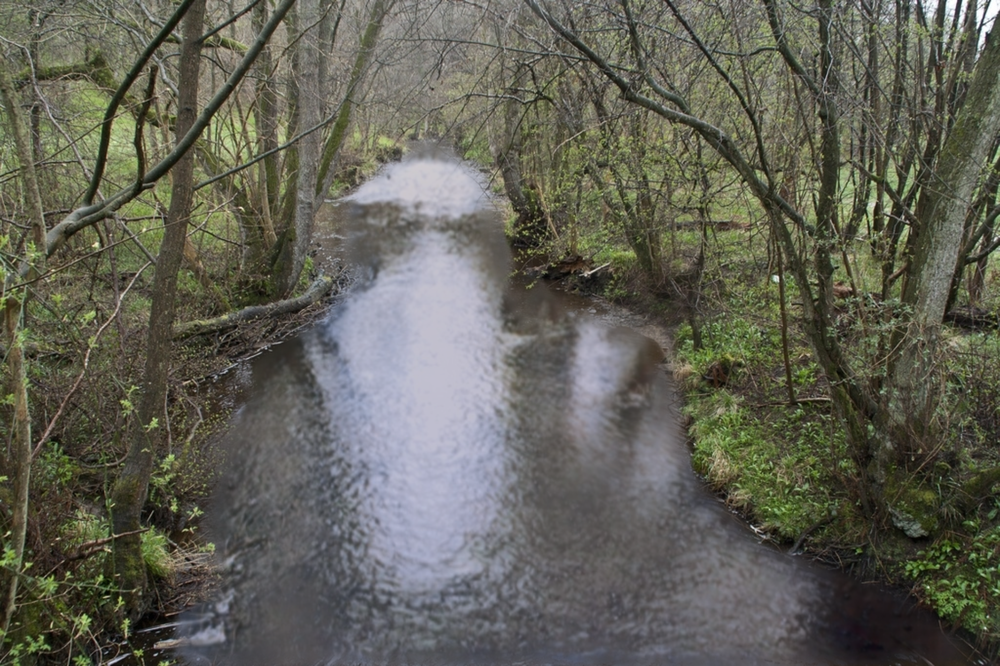}
         \includegraphics[keepaspectratio, width=\hsize]{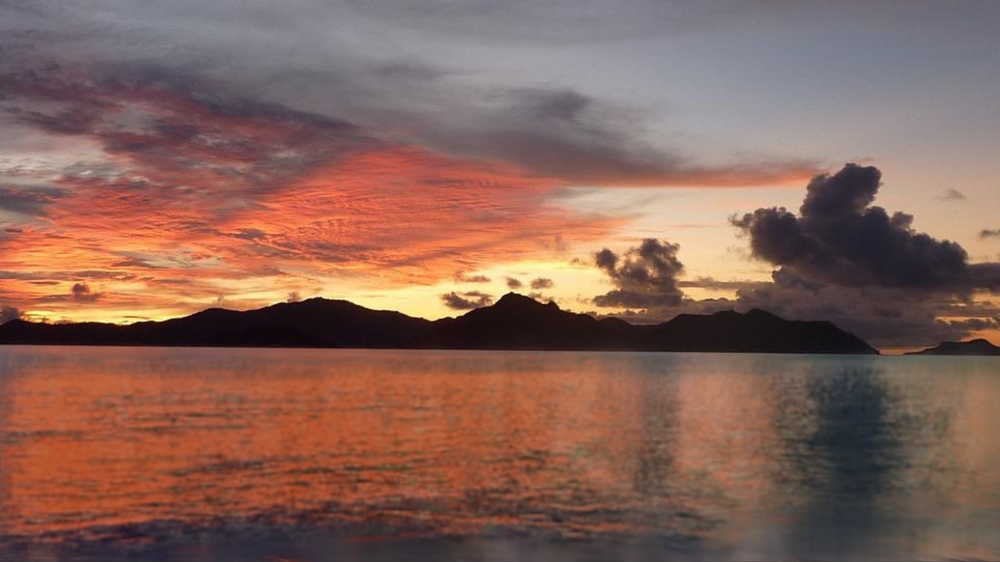}
         \includegraphics[keepaspectratio, width=\hsize]{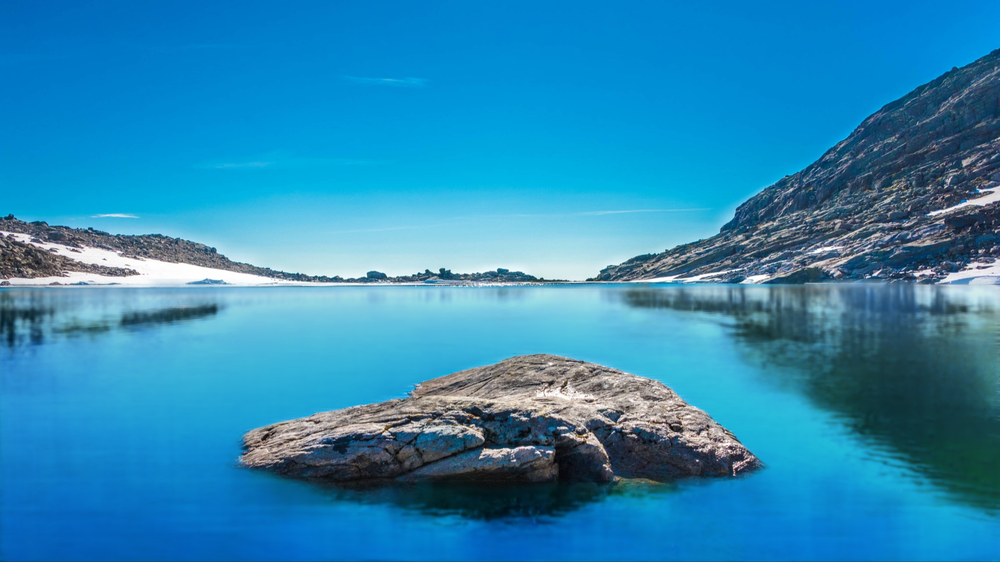}
        \includegraphics[keepaspectratio, width=\hsize]{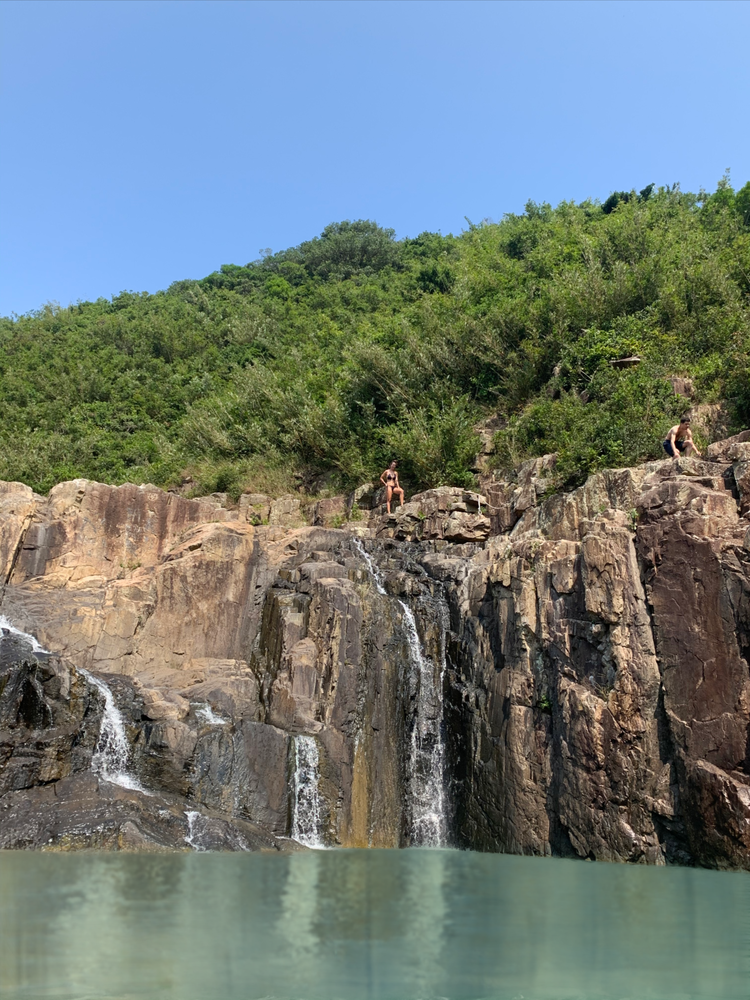}
        \includegraphics[keepaspectratio, width=\hsize]{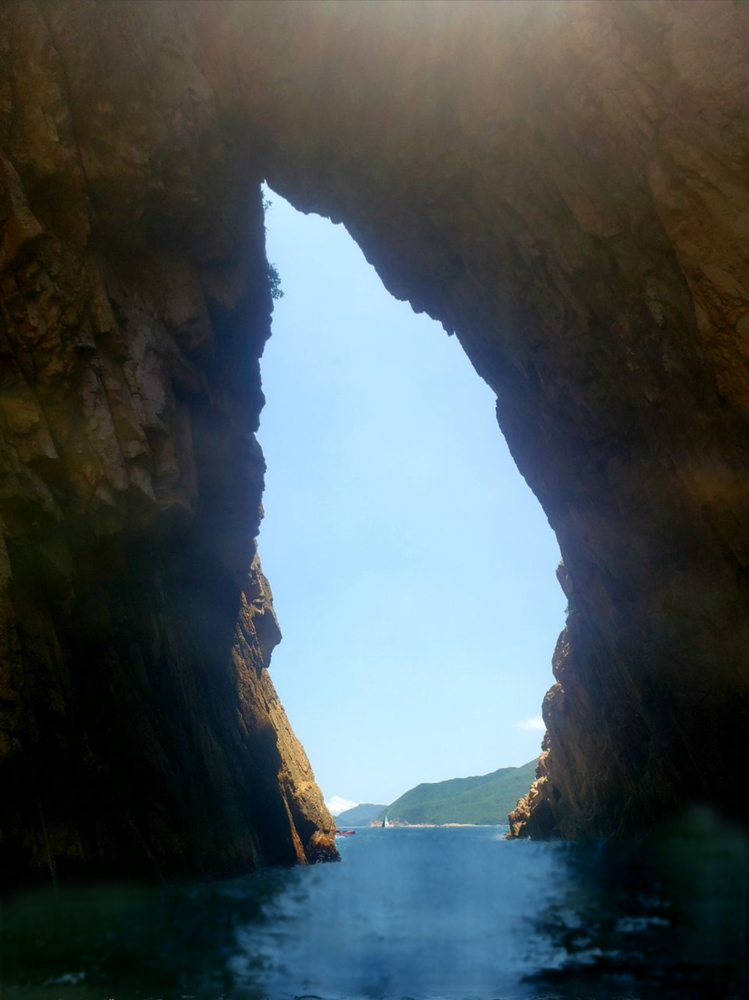}
    \end{minipage}
    \begin{minipage}{0.205\hsize}
        \centering
        Input
    \end{minipage}
    \begin{minipage}{0.205\hsize}
        \centering
        DISTS + Color
    \end{minipage}
    \begin{minipage}{0.205\hsize}
        \centering
        DISTS Only
    \end{minipage}
    \begin{minipage}{0.205\hsize}
        \centering
        L1
    \end{minipage}
    \caption{Comparison with parameters estimated with different energy functions for the cuckoo search. From left, inputs, results with the combination of DISTS and the color metric, results with DISTS metric only, and results with L1 loss. The results are obtained with 500 iterations for consistency. We obtain the best results using the combination of DISTS and the color metric. See the main paper for the water masks and the reflection textures used. The input images: - Paul -/Flickr (first), Edward Nicholl/Flickr (second), Foliez/Pixabay (third), and Unknown/PxHere (fourth).}
    \label{fig:ablation}
\end{figure*}

In the design of the energy function, we chose the DISTS metric because it is not sensitive to local variations caused by wave dynamics, as opposed to other simpler image metrics. In Fig.~\ref{fig:ablation}, when the L1 loss is used as the energy function instead of DISTS, it fails to estimate the water dynamics completely, generating waves with either very high or very low frequency components only. These are expected results because we do not expect the wave dynamics of the input image and the rendered results to match exactly, and simple pixel-wise metrics cannot compute the perceptual difference between the two images. 
In addition to DISTS, the color similarity metric is employed to further penalize the color dissimilarity. In Fig.~\ref{fig:ablation}, we can observe that the addition of color dissimilarity energy lets the optimized parameters generate more visually plausible results by correcting the overall color shifts that are otherwise present in some results.
We limit our discussion to the qualitative evaluation in this work because a quantitative evaluation would be difficult because of the local variations of water dynamics.

\section{Rendering optimizations}
The basic method for computing the reflection color described in the paper is computationally inefficient mainly due to the ray marching step. We can reduce the number of iterations by using larger ray marching step sizes, but that can significantly degrade rendering quality. We observe that the collision point for the same reflection vector only changes when the viewpoint moves (user zooms or pans). Based on these observations, we can first precompute the ray marching step and reuse the results for multiple frames while the viewpoint is static.

Furthermore, in the precomputation step, we partition pixels in $4 \times 4$ groups in image space. For each group of 16 pixels, we precompute the potential collision points in screen space for 16 reflection directions uniformly distributed in the hemisphere, storing the results in a texture. In our implementation, we store at most two collision points (first and last) for each direction within each pixel group. The rationale is that the last collision should be against the same common ``background'' wall, while the first ray collision may be against different smaller walls. Finally, in the rendering pass, we retrieve the screen space collision points by interpolating the precomputed collision points in screen space instead of computing anew. 

These optimizations increase the performance by approximately 10$\times$, yielding accurate results in the vast majority of cases, but occasionally producing artifacts in complex boundary regions (Fig.~\ref{fig:renderer_optim_success}). All results in the paper and supplemental material use these optimizations.

\begin{figure}[t]
    \centering
    \includegraphics[keepaspectratio, width=\hsize]{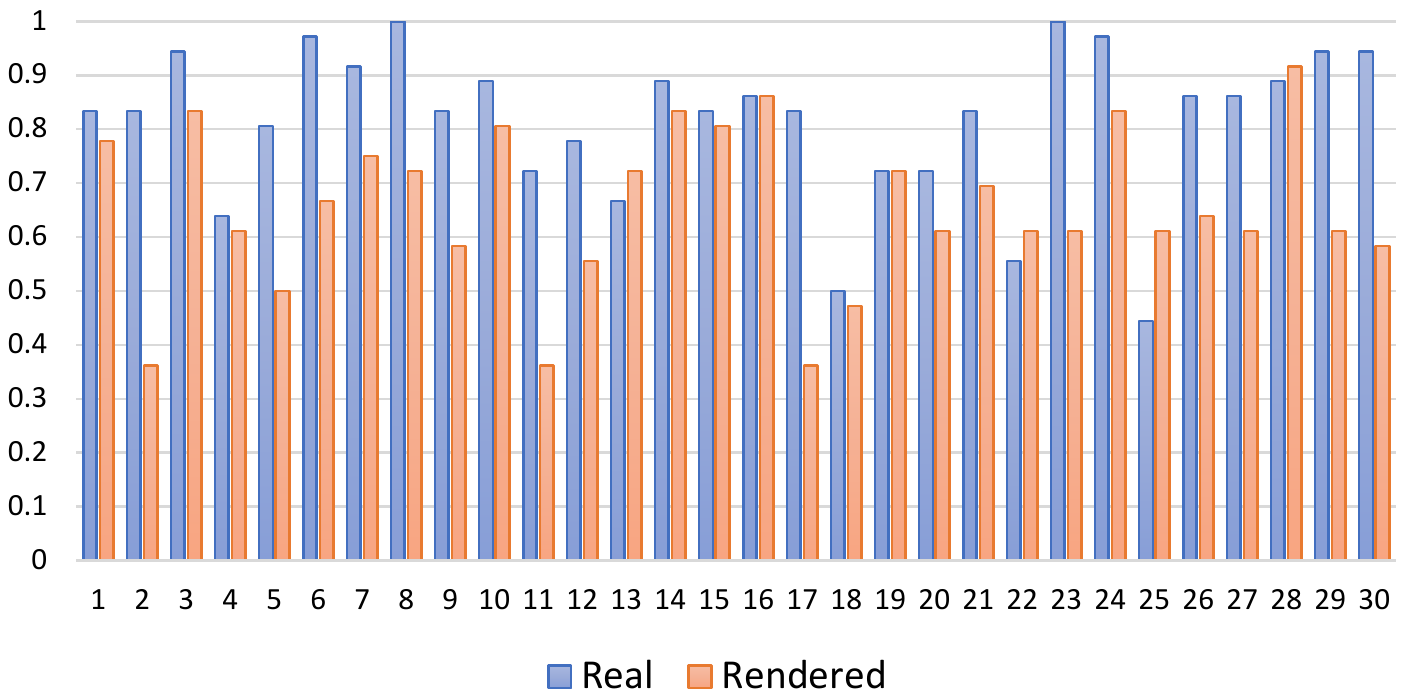} \\
    (a) statistics \\
    \includegraphics[keepaspectratio, width=\hsize]{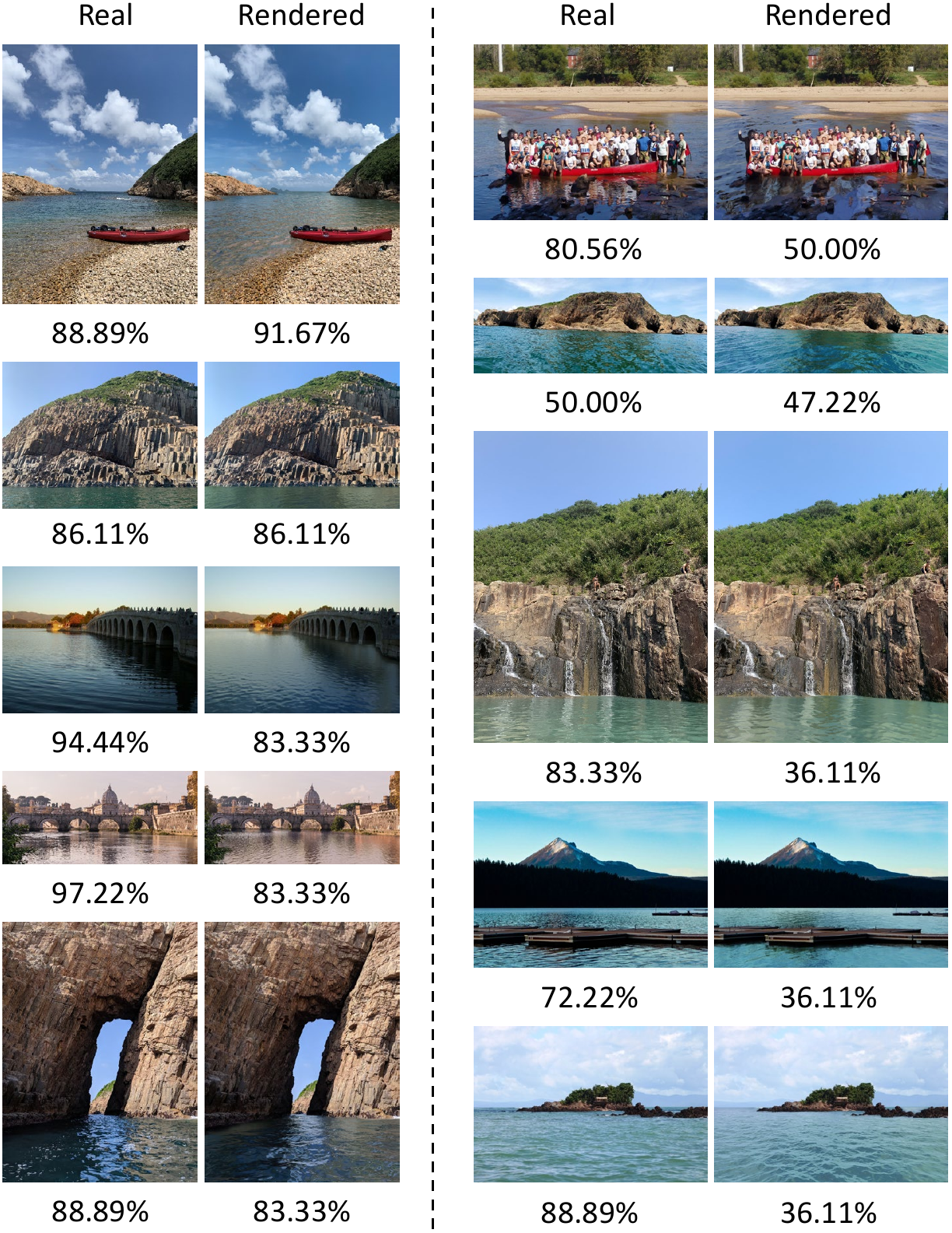} \\ 
    (b) examples \\
    \caption{User study results. (a) illustrates the percentage of real/rendered images selected as real in every case; and (b) demonstrates the top 5 cases where the rendered images are thought as most photorealistic (left) and the top 5 cases where the rendered images are selected as least photorealistic (right). In (b), the numbers indicate the percantages of real/rendered images selected as real.}
    \label{fig:user_study_cases}
\end{figure}

\section{Details of user study}
The user study is designed to better evaluate our photorealistic synthesis results despite the difference in their water dynamics compared to the input images. So the test cases in the user study mix real images and synthesis results, which are then shown to users in random order. 
More specifically, we first randomly selected 30 input images from the testing dataset (which has 67 images in total) and then generated 30 test cases for every user. In each case, either the input image or its rendered result was shown to the user, who was asked to determined whether it was a photo or synthetic result. Users were given unlimited time to view the image and make a choice. A total of 72 users participated and 2160 answers were collected. The same number of real and rendered images were shown. The result shows that users successfully identified these real images as real 81.67\% of the time, indicating that the users understood the task, while the rendered images fooled users in 65.46\% of cases, which shows that most of our generated results are thought as realistic as real images. Fig.~\ref{fig:user_study_cases} (a) shows the percentage of real/render image selected as real in each case, and Fig.~\ref{fig:user_study_cases} (b) demonstrates five most photorealistic and five least photorealistic rendered images voted by users. Note that our method is able to generate photorealistic results even though the predicted water animation may not exactly match the input image. 

\bibliographystyle{ACM-Reference-Format}
\bibliography{main}